\documentclass[aps,twocolumn,showpacs,superscriptaddress,nofootinbib]{revtex4-2}

\usepackage{graphicx}
\usepackage{latexsym}
\usepackage{amsmath}

\usepackage{color}
\newcommand{\red}[1]{\textcolor[named]{Red}{#1}}
\newcommand{\blue}[1]{\textcolor[named]{Blue}{#1}}

\begin{document}
\title{
Langevin analogy between particle trajectories and polymer configurations
}

\author{Takuya Saito}
\email{tsaito@phys.aoyama.ac.jp}
\affiliation{Department of Physical Sciences, Aoyama Gakuin University, Chuo-ku, Sagamihara 252-5258, Japan}

\def\Vec#1{\mbox{\boldmath $#1$}}
\def\degC{\kern-.2em\r{}\kern-.3em C}

\def\SimIneA{\hspace{0.3em}\raisebox{0.4ex}{$<$}\hspace{-0.75em}\raisebox{-.7ex}{$\sim$}\hspace{0.3em}} 

\def\SimIneB{\hspace{0.3em}\raisebox{0.4ex}{$>$}\hspace{-0.75em}\raisebox{-.7ex}{$\sim$}\hspace{0.3em}}

\date{\today}

\begin{abstract}
A diffusive trajectory drawn by the generalized Langevin equation (GLE) for a colloidal particle evokes a random fractal of a static polymer configuration.
This article proposes a static GLE-like description that enables the generation of a single configuration of a polymer chain with the noise formulated to satisfy the static fluctuation-response relation (FRR) along a one-dimensional chain structure but not along a temporal coordinate.
A remarkable point is qualitative differences and similarities in the FRR formulation between the static and the dynamical GLEs.
Guided by the static FRR, we further make analogous arguments in the light of stochastic energetics and the steady-state fluctuation theorem.
\end{abstract}


\def\degC{\kern-.2em\r{}\kern-.3em C}

\newcommand{\gsim}{\hspace{0.3em}\raisebox{0.5ex}{$>$}\hspace{-0.75em}\raisebox{-.7ex}{$\sim$}\hspace{0.3em}} 
\newcommand{\lsim}{\hspace{0.3em}\raisebox{0.5ex}{$<$}\hspace{-0.75em}\raisebox{-.7ex}{$\sim$}\hspace{0.3em}} 

\def\Vec#1{\mbox{\boldmath $#1$}}

\maketitle

\section{Introduction}
\label{Intro}

A close observation of stochastic phenomena ubiquitously reveals underlying fractal structures~\cite{Mandelbrot,PhysRep_Metzler_2000,AdvPhys_Havlin_2002,JPhys_Forrest_Witten_1979,PRL_Witten_Sander_1981,PhysRep_Bouchaud_Georges_1990,JPSJ_Matsushita_1991}.
Not only nonequilibrium but also equilibrium conditions create the fractal figures.
A typical example is polymer configurations, whose snapshots display random fractals due to chain connectivity in thermal agitation~\cite{deGennesBook,Doi_Edwards,Grosberg_Khokhlov,PhysRep_Vilgis_2000}.
Indeed, two-point correlations grow as a power law:
\begin{eqnarray}
\left< \Delta x_n^2 \right> \sim n^{2\nu},
\label{x_static}
\end{eqnarray}
where $x_n$ indicates the $n$-th monomer position with a monomer index $n$ along a polymer chain  and $\left< \cdot \right>$ is considered an appropriate average with $\Delta x_n=x_n-x_0$.
The Flory exponent $\nu$ characterizes the spatial fluctuation size of a polymer (e.g., $\nu=1/2$ for an ideal chain, whereas $\nu=3/4$ or $\simeq 0.588$ for a self-avoiding (SA) chain in two or three dimensions, respectively)~\cite{deGennesBook,Doi_Edwards,Grosberg_Khokhlov,PhysRep_Vilgis_2000}. 
The inverse exponent $1/\nu$ describes the fractal dimension.

In addition to the spatial configurations, a rich variety of stochastic processes are also subject to a fractal notion, as observed by inspecting trajectories plotted as a function of time~\cite{Mandelbrot,PhysRep_Metzler_2000,AdvPhys_Havlin_2002,PhysToday_Barkai_2012,vanKampen,Gardiner}. The spatial correlation (eq.~(\ref{x_static})) is replaced by the mean square displacements (MSDs):
\begin{eqnarray}
\left< \Delta x(t)^2 \right> \sim t^{\alpha},
\label{MSD_x}
\end{eqnarray}
where $\Delta x(t)=x(t)-x(0)$ with time $t$.
The Brownian motion or random walk with $\alpha=1$ offers a primary illustration referred to as normal diffusion.
In addition, the notion is extended to anomalous diffusion identified as nonlinear growth with $\alpha\neq 1$~\cite{Mandelbrot,PhysRep_Metzler_2000,AdvPhys_Havlin_2002,NewJPhys_Godec_2014,PRE_Lizana_Barkai_Lomholt_2010,PhysToday_Barkai_2012,Amitai,Krug,BRL_Ooshida_2016,JCP_Schiessel_1995,JStatMech_Panja_2010,PRE_Sakaue_2013,PRE_Saito_2015,TMP_Burlatskii_Oshanin_1988,SoftMatter_Ernst_Weiss_2012,PRE_Weiss_2013}.
Interestingly, the static configurations for the ideal or SA chain have a mathematical analogy to the trajectories drawn by the random or SA walk, respectively~\cite{deGennesBook,Doi_Edwards,Grosberg_Khokhlov,PhysRep_Vilgis_2000}.

Consider the dynamical side of the polymer.
The temporal evolution of a tagged monomer's position $x(t)$ belongs to a class of anomalous diffusion expressed by the generalized Langevin equation (GLE) with a power-law-function memory kernel $\mu(t)$~\cite{JCP_Schiessel_1995,JStatMech_Panja_2010,PRE_Sakaue_2013,PRE_Saito_2015}:
\begin{eqnarray}
\frac{dx(t)}{dt}
=\int_0^tds\, \mu(t-s) f(s) +\eta(t),
\label{GLE_x}
\end{eqnarray}
where $f(t)$ or $\eta(t)$ denotes an externally controlled force or noise acting on the tagged monomer, respectively.
A striking similarity between the statics and the dynamics in the polymer leads to an obvious question:
Can the stochastic differential equation akin to GLE~(\ref{GLE_x}) describe a static polymer configuration with an altered kernel $\mu_{st}(n,n')$?
\begin{eqnarray}
\frac{dx_n}{dn}
=\int_0^N dn'\, \mu_{st}(n,n')f_{n'} +\overline{\eta}_n,
\label{GLE_xn}
\end{eqnarray}
where $f_{n'}$ is an externally controlled constant force acting on the $n'$-th monomer and $\overline{\eta}_n$ is static noise.
An analogous formulation underlying the polymer is, however, elusive.

Our aim in this article is to find a stochastic analogy between the particle trajectories and the polymer configurations.
In section~\ref{GLEanalogy}, we begin with a review of the conventional GLE derived from the mode analyses and then propose a static GLE-like description with the static FRR not along time $t$ but along the monomer index $n$.
Section~\ref{micro_der}  
addresses the same statistical issue to relate a static response with fluctuations (the static FRR)
from a partition function based on a path integral.
Once the Langevin analogy is found, the stochastic energetics~\cite{SekimotoBook} or the fluctuation theorem~\cite{RPP_Seifert_2012,PRE_Crooks_1999,JPhysA_Kurchan_1998,JStatPhys_Lebowitz_Spohn_1999,JCP_Gaspard_2004}, which has been developed by focusing on a particle trajectory, may fall within the scope of the analogy to a polymer configuration. 
In section~\ref{SE_analogy}, we consider the heat analogy to the stochastic energetics proposed by Sekimoto~\cite{SekimotoBook} with the Langevin equation. 
To be consistent with the heat analogy, we develop the argument toward the steady-state-fluctuation-theorem analogy in section~\ref{NSS}.
In section~\ref{discussion}, we discuss applications and perspectives.
Section~\ref{conclusion} concludes this study.

\section{Langevin analogy}
\label{GLEanalogy}

\subsection{From a ``dynamical" Rouse chain to a ``static" ideal chain} 
{\it --- Dynamics ---} 
We first review a derivation of the dynamical GLE from equations of motion for the Rouse model~\cite{JStatMech_Panja_2010}.
Consider a linear polymer consisting of $N$ monomers in a viscous solution at temperature $T$.
Also consider monomers indexed from one chain end by $n$ and whose position at time $t$ is expressed as $x_n(t)$. 
Note that we focus on a one-dimensional coordinate along a forced direction, on which the polymer motion is projected.
The equation of motion for each monomer is written as
\begin{eqnarray}
\gamma \frac{\partial x_n(t)}{\partial t}
=
k\frac{\partial^2 x_n(t)}{\partial n^2}
+f_n(t)+\zeta_n(t),
\label{eq_motion}
\end{eqnarray}
where $\gamma$ denotes the frictional coefficient per monomer and $k$ is the spring constant.
On the right-hand side, the last two terms $f_n(t)$ and $\zeta_n(t)$ represent the external force and the thermal noise acting on monomer $n$, respectively.
The noise is the mean zero $\left< \zeta_n(t)\right> =0$ and the Gaussian distributed random force with the correlation $\left< \zeta_n(t)\zeta_{n'}(t) \right> = 2N\gamma k_BT\delta (t-t')\delta (n-n')$.\footnote{A conversion from discrete to continuous forms implies that $\delta_{nn'}/N \simeq \delta (n-n')$ for a sufficiently large $N$.}
In the noisy diffusion equation (\ref{eq_motion}), the mean force balance $\gamma \partial \left<x_n(t)\right>/\partial t\simeq k \partial^2 \left<x_n(t)\right>/\partial n^2$ represents tension propagation along the chain backbone~\cite{PRE_Sakaue_2012}, whereas $\zeta_n(t)$ plays a role in diffusion over the spatial dimension.

We now trace $x_n(t)$ at the controlled force with $f_n=f \delta (n-n_0) \Theta (t)$, where the force is applied into the $n_0$-th monomer,\footnote{For the external force acting on or around the chain end $n_0=0$ or $N$, a positive infinitesimal $\epsilon$ is added or subtracted such that $n_0=\epsilon$ or $n_0=N-\epsilon$, respectively, for technical reasons.} and externally controlled force with $f$ or $\Theta (t)$ being the force magnitude or the Heaviside step function, respectively~\cite{JStatMech_Panja_2010,PRE_Sakaue_2013,PRE_Saito_2015,PRE_Saito_Sakaue_2017,PRE_Saito_2017,Polymers_Saito_Sakaue_2019}.

To deal with eq.~(\ref{eq_motion}), we move to the normal mode $X_q(t)$~\cite{Doi_Edwards}:
\begin{eqnarray}
\gamma_q \frac{\partial X_q(t)}{\partial t}
=
-k_q X_q
+F_q(t)+Z_q(t)
\label{x_mode}
\end{eqnarray}
with the conversion between the real and mode space defined as
\begin{eqnarray}
X_q(t)=\int_0^N dn\,x_n(t)h_{q,n}, \quad x_n(t)=\sum_{q=0}^N X_q(t)h_{q,n}^\dagger.
\label{Xq}
\end{eqnarray}
The conversion kernels in eqs.~(\ref{Xq}) are introduced as
\begin{eqnarray}
h_{q,n} =\frac{1}{N}\cos{\left( \frac{\pi qn}{N} \right)}, \quad 
h_{q,n}^\dagger = \frac{1}{c_q} \cos{\left( \frac{\pi qn}{N} \right)}
\label{h_qn}
\end{eqnarray}
to satisfy a free boundary condition $\partial x_n/\partial n|_{n=0}=\partial x_n/\partial n|_{n=N}=0$~\cite{Doi_Edwards} with $c_q\equiv (1+\delta_{q0})/2$.
The coefficients for the Rouse model are assigned with
\begin{eqnarray}
k_q=k(\pi q/N)^{2}, \quad \gamma_q=\gamma.
\label{Rouse_coeff}
\end{eqnarray}
As with eq.~(\ref{Xq}), $f_n(t)$ or $\zeta_n(t)$ becomes $F_q(t)$ or $Z_q(t)$ on the mode space, respectively.
The noise $Z_q(t)$ in the normal coordinate has the mean zero $\left< Z_q(t)\right>=0$ and satisfies the equilibrium noise condition:
\begin{eqnarray}
\left< Z_q(t)Z_{q'}(t') \right>= \frac{2c_qk_BT\gamma_q}{N}\delta (t-t')\delta_{qq'},
\label{FDR_x}
\end{eqnarray}
which implies that, in practice, modal motion for $q$ is subject to the thermal bath with effective temperature $c_qT/N$.

Superposition of the normal modes gives a solution to eq.~(\ref{eq_motion}):
\begin{eqnarray}
x_n(t)
&=&
\sum_q\int_{-\infty}^tds\,\frac{F_q(s)+Z_q(s)}{\gamma_q} e^{-(t-s)(k_q/\gamma_q)}h_{q,n}^\dagger.
\label{solution_x}
\end{eqnarray}
If only the $n$-th monomer is traced, the subscript drops as $x_n(t) \rightarrow x(t)$, which is hereinafter called a tagged monomer.
The tagged monomer dynamics is found to accompany the anomalous diffusion generated by the GLE combined with power-law memory kernels~\cite{JStatMech_Panja_2010,PRE_Sakaue_2013,PRE_Saito_2015,PRE_Saito_Sakaue_2017,PRE_Saito_2017}, where the external force is given by $f(t)=f\Theta (t)$ and $\eta(t)$ is Gaussian distributed noise with mean zero $\left< \eta(t)\right>=0$ and covariance
\begin{eqnarray}
\left< \eta(t)\eta(s) \right>=k_BT\mu(t-s).
\label{GLE_x_FDR}
\end{eqnarray}
Equation~(\ref{GLE_x_FDR}) is referred to as the fluctuation-response relation (FRR) and associates the equilibrium noise with the response function for the GLE.

Note that, to be exact, we define $\mu(s-t)\equiv \mu(t-s)$ for $s\geq t$ using the mobility kernel $\mu(t-s)$ for $t\geq s$, whereas the mobility kernels that account for causality are expressed as $\delta\left< dx(t)/dt\right>/\delta f(s)=\mu(t-s)\Theta(t-s)$.
A response to the external stimulus is identified by the mobility kernel:
\begin{eqnarray}
\mu(t) \simeq \mu_{c.m.}(t) + \mu_{0}(t) +\mu_{\alpha}(t),
\label{GLE_memory_x}
\end{eqnarray}
which is broken down into three components verified by close inspection of the time derivative of eq.~(\ref{solution_x}).
From left to right on the right-hand side, the respective terms express the center of mass motion $\mu_{c.m.}(t)\simeq 2\delta (t)/(N\gamma_1)$, the instantaneous monomeric response $\mu_{0}(t)\simeq 2\gamma^{-1}\delta (t)$ before detecting the chain connectivity, and a relaxation of the internal configuration $\mu_\alpha(t)\sim |t|^{\alpha-2}$ with $\alpha=1/2$ for the Rouse polymer.

In eq.~(\ref{GLE_memory_x}), the internal relaxation $\mu_\alpha(t)$ that generates the subdiffusion ($\alpha<1$) is noteworthy. Monitoring the internal relaxation regime during the longest-relaxation-time period $\tau_1=\gamma_1/k_1=N^2 (\gamma/k)$, we encounter the subdiffusion verified with $\mu_\alpha (t)$ as $\left< \Delta x(t)^2 \right> = \int_0^tds\int_0^tds'\, \left< \eta(s)\eta(s') \right> \simeq k_BT\int_0^tds\int_0^tds'\, \mu_\alpha (s-s') \sim t^{1/2}$.

{\it --- statics ---} 
We here attempt to find the static GLE-like expression that describes each configuration of the ideal chain\footnote{Conventionally, the Rouse model assumes the basic equation to discuss the dynamics with the local friction, whereas the ideal chain is the static representation without specification of the dissipation mechanism.} as an analogy for eq.~(\ref{GLE_x}).
One may naturally expect eq.~(\ref{GLE_xn}).
Note that the mean gradient of tension is reduced to the applied force $f_n\equiv -dT_n/dn$, where $T_n$ is referred to as the ``applied tension" acting on the $n$-th monomer with $\int_0^Ndn\,dT_n/dn=T_N-T_0=0$ taken as a whole.
Notably, the tension $T_n$ behaves like a ``force potential" invoked by an ``energy potential," whereas the respective potentials provide the force from a derivative with respect to ``$n$" or ``$x$".
In addition, $\mu_{st}(n)$ denotes a static (superdiffusive) kernel, and $\overline{\eta}_n$ is static noise with mean zero $\left< \overline{\eta}_n\right>=0$.

The relation of the static kernel $\mu_{st}(n)$ with $\overline{\eta}_n$ no longer shares the same form as eqs.~(\ref{GLE_x_FDR}); we now derive an analogous static FRR using the solution eq.~(\ref{solution_x}).

The derivative of $x_n(t)=\sum_{q \geq 1} X_q(t)h_{q,n}^\dagger$ with respect to $n$ produces the left side of eq.~(\ref{GLE_xn}):
\begin{eqnarray}
\frac{dx_n(t)}{dn}
&=&
-\sum_q\int_{-\infty}^tds\,\frac{\pi q}{N}\frac{F_q+Z_q(s)}{\gamma_q} 
\nonumber \\
&&
\times e^{-(t-s)(k_q/\gamma_q)}h_{q,n}^{(s)\dagger},
\label{dxn_dn_1}
\end{eqnarray}
where kernels distinct from $h_{q,n}$, $h_{q,n}^\dagger$ are introduced as 
\begin{eqnarray}
h_{q,n}^{(s)} =\frac{1}{N}\sin{\left( \frac{\pi qn}{N} \right)}, \quad 
h_{q,n}^{(s)\dagger} = \frac{1}{c_q} \sin{\left( \frac{\pi qn}{N} \right)}.
\label{h_qn_sin}
\end{eqnarray}
The superscript $(s)$ is used to remind us of the initial letter of $\sin{(\cdot)}$.
Note that, in practice, $\int_0^Ndn\,dT_n/dn=0$ does not demand that $q=0$ mode be considered. 
The right-hand side of eq.~(\ref{dxn_dn_1}) should be reduced to the right-hand side of eq.~(\ref{GLE_xn}).
When the deterministic and the stochastic parts are separated, the static fluctuation $\overline{\eta}_n$ turns out to be given by
\begin{eqnarray}
\overline{\eta}_n
&=&
-\sum_q\frac{\pi q}{N} 
\left( 
\int_{-\infty}^{t} ds\, \frac{Z_q(s)}{\gamma_q} e^{-(t-s)(k_q/\gamma_q)}
\right)
h_{q,n}^{(s)\dagger}
\label{noise_xn_Xq}
\end{eqnarray}
Using eq.~(\ref{noise_xn_Xq}), we calculate the autocorrelation of noise, which is at equal time between $n$ and $n'$:\footnote{
The external force $f_m$ determines gradient of tension $f_m=-dT_m/dm$ and $F_q=\int_0^Ndm\,f_mh_{q,m}$.
The mean part of eq.~(\ref{dxn_dn_1}) is extracted as
\begin{eqnarray}
&&-\sum_q\int_{-\infty}^tds\,\frac{\pi q}{N}\frac{F_q}{\gamma_q} e^{-(t-s)(k_q/\gamma_q)}h_{q,n}^{(s)\dagger}
\nonumber \\
&=&
\red{-}\sum_q \frac{1}{k_q}
\frac{\pi q}{N}F_qh_{q,n}^{(s)\dagger} 
\nonumber \\
&=&
-\sum_{q,q',m} \frac{\pi q}{N} \frac{1}{k_q}h_{q,n}^{(s)\dagger}h_{q,m}
F_{q'}h_{q',m}^{\dagger} 
\nonumber \\
&=&
\sum_{m} \left(-\sum_{q} \frac{\pi q}{N} \frac{1}{k_q} h_{q,n}^{(s)\dagger}h_{q,m} \right)
\left( \sum_{q'} F_{q'}h_{q',m}^{\dagger} \right)
\label{mode_mu_F}
\end{eqnarray}
To find eq.~(\ref{mu_st}), we compare eq.~(\ref{mode_mu_F}) with eq.~(\ref{GLE_xn}) and also use $\sum_m h_{q,m}^{(s)}h_{q',m}^{(s)\dagger}=\delta_{qq'}$.
}
\begin{eqnarray}
\left< \overline{\eta}_n\overline{\eta}_{n'} \right> 
&=&
\sum_{q\geq 1}
\left( \frac{\pi q}{N} \right)^2
\int_{-\infty}^t ds\, \frac{2c_qk_BT}{N\gamma_q} 
\nonumber \\
&&
\times
e^{-2(t-s)(k_q/\gamma_q)}
h_{q,n}^{(s)\dagger}h_{q,n'}^{(s)\dagger}
\nonumber \\
&=&
\sum_{q\geq 1}
\left( \frac{\pi q}{N} \right)^2
\frac{c_qk_BT}{Nk_q} 
h_{q,n}^{(s)\dagger}h_{q,n'}^{(s)\dagger}.
\label{noise_xn_corr}
\end{eqnarray}
A remaining issue is $\mu_{st}(n)$.
By superimposing the mean mode components, we find the static memory kernel to be
\begin{eqnarray}
\mu_{st}(n,n') 
&=& \red{-} \sum_{q \geq 1} \frac{1}{k_q} \left( \frac{\pi q}{N} \right) h_{q,n'}h_{q,n}^{(s)^\dagger}.
\label{mu_st}
\end{eqnarray}
Comparing eq.~(\ref{mu_st}) with (\ref{noise_xn_corr}) reveals that a simple replacement from $t$ to $n$ does not successfully establish the FRR like eq.~(\ref{GLE_x_FDR}).
Instead, we encounter
\begin{eqnarray}
\left< \overline{\eta}_n \overline{\eta}_{n'} \right> 
= k_BT \partial_{n'} \mu_{st}(n,n'),
\label{static_FDR_u}
\end{eqnarray}
where $\partial_{n'} h_{q,n'}=-(\pi q/N) (c_q/N) h_{q,n'}^{(s)^\dagger}$ with a notion $\partial_n \equiv d/dn$.
Notably, eq.~(\ref{static_FDR_u}) is specified only by the static parameter $k_q$, similar to a static relation between heat capacity and internal energy fluctuation $k_BT^2 \partial \left< U\right>/\partial T=\left< \delta U^2 \right>$, where $U$ denotes the internal energy.
In addition, we find that, generally, $\mu_{st}(n,n')\neq \mu_{st}(n',n)$; however, symmetricity is recovered by differentiating it as $\partial_{n'} \mu_{st}(n,n')=\partial_{n} \mu_{st}(n',n)$.

\begin{figure}[t]
\begin{center}
\includegraphics[scale=0.5]{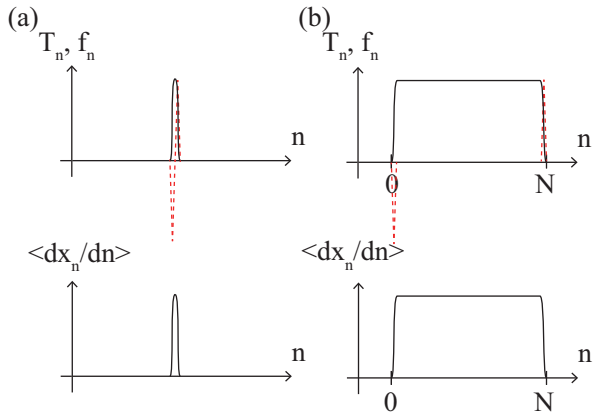}
      \caption{
		(Color online) Schematic profiles for a Rouse polymer in tension $T_n$, applied force $f_n$, and local stretching rate $\left< dx_n/dn \right>$:
		(a) local tension induced by a local external dipole force and (b) uniform tension applied by pulling both the chain ends.
		The upper figures show profiles for tension or applied force, drawn with a solid black or red broken line, respectively.
		The corresponding local stretching rate $\left< dx_n/dn \right>$ is displayed in the lower figures.
		The vertical scales for $T_n$ and $f_n$ are not the same.
	  }
\label{sketch_fig_Rouse}
\end{center}
\end{figure}

We here consider specific situations.
Only the local interaction is present for the Rouse polymer with $\nu^{(x)}=1/2$, where the noise serves as a local point-like correlation:
\begin{eqnarray}
\left< \overline{\eta}_n \overline{\eta}_{n'} \right> =\frac{k_BT}{k}\delta (n-n'),
\label{Rouse_static_noise_corr}
\end{eqnarray}
which is derived from eq.~(\ref{noise_xn_corr}) as the continuum limit of the monomer index.
Then, recalling eq.~(\ref{static_FDR_u}) and $d\left<x_n(t)\right>/dn=-\int dn'\, \mu_{st}(n,n')dT_{n'}/dn'$, we estimate the mean stretching rate $d\left<x_n(t)\right>/dn$.
Specifically, if a local dipole force with $f_{n}=-\partial T_n/\partial n=-f\partial \delta (n-n_d)/\partial n$ acts on $n_d$-th monomer, the mean rate is found to be $d\left<x_n\right>/dn=(f/k) \delta (n-n_d)$, indicating local stretching at $n_d$ (see fig.~\ref{sketch_fig_Rouse}\,(a)).
As shown in the profiles for the tension (black solid line) and the applied force (red broken line) in fig.~\ref{sketch_fig_Rouse}\,(a), a pair of applied external forces with equivalent magnitude but opposite directions (i.e., the dipole expressed mathematically by $f_{n}=-f\partial \delta (n-n_d)/\partial n$) acts if the tension $T_{n}=f\delta (n-n_d)$ is locally imposed there,\footnote{
A method for handling the delta function is found in a textbook on qualitative seismology~\cite{Aki_Richards_Book} and also in a paper on chromatin hydrodynamics~\cite{BiophysJ_Bruinsma_Grosberg_2014}.
} 
which stretches the local part like the delta function, as shown in the lower-left profile in fig.~\ref{sketch_fig_Rouse}.
In addition, as shown in fig.~\ref{sketch_fig_Rouse}\,(b), if both the ends are pulled with $T_n=f\Theta (n-\epsilon)-f\Theta (n-N+\epsilon)$, we observe homogeneous stretching $d\left< x_n\right>/dn=f/k$ as a result of the local restoring force expressed by $k\partial^2 x_n/\partial n^2$ under uniform tension.
Note that, for technical purposes, positive infinitesimal $\epsilon$ is introduced such that the forced points are completely in the chain.\footnote{
Both the Rouse and SA polymers can use eq.~(\ref{mean_stretching}) to derive the local stretching rate $d\left< x_n\right>/dn$.
Taking the average of eq.~(\ref{GLE_xn}), we have
\begin{eqnarray}
\frac{d\left< x_n\right>}{dn}
&=&
-
\int_0^N dn'\, \mu_{st}(n,n')\frac{dT_{n'}}{dn'}
\nonumber \\
&=&
\int_0^N dn'\, \frac{\partial \mu_{st}(n,n')}{\partial n'}T_{n'}
=
(k_BT)^{-1}
\int_0^N dn'\, 
\left< \overline{\eta}_n \overline{\eta}_{n'} \right>T_{n'}
\nonumber \\
\label{mean_stretching}
\end{eqnarray}
The second equation is followed by integration by parts with $T_0=T_N=0$, and the last equation comes from the FRR eq.~(\ref{static_FDR_u}).
}

\subsection{SA chain}

{\it --- SA polymer ---} Often, in practical situations, incorporating long-range interactions is inevitable.
First, the monomers repel each other and the effective interaction of SA qualitatively swells the polymer, as quantified by the Flory exponent $\nu$ mentioned in the discussion of eq.~(\ref{x_static}).
In addition, together with the SA interactions, we discuss the hydrodynamic interactions (HIs) that lead to qualitative changes in the dynamics, although the HIs are not necessarily required in the SA chain.
The HIs add long-range frictional interaction between monomers due to the medium flow created by the distant monomer's motion.
The characteristic relaxation time is represented by $\tau \sim R^z$ with dynamical exponent $z$, e.g., $z=3$ for the nondraining scenario or $z=2+1/\nu$ for the free-draining scenario. A generalization that incorporates these long-range interactions into mode analyses\footnote{
The numerical verification in the mode analyses of the SA interaction is found in ref.~\cite{JCP_Panja_2009}.
}
can be implemented by replacing the coefficients in eqs.~(\ref{x_mode}),\,(\ref{FDR_x}) with
\begin{eqnarray}
k_q=k(q/N)^{1+2\nu}, \quad \gamma_q=\gamma(q/N)^{1-(z-2)\nu}.
\label{gen_coeff}
\end{eqnarray}
Incidentally, eq.~(\ref{Rouse_coeff}) for the Rouse polymer with the local interaction is recovered by substituting $\nu=1/2$ and $z=4$. Likewise, the GLE form of eq.~(\ref{GLE_x}) can be constructed by superposition of the whole modes: $x(t)=\sum_q X_q(t)h_{q,n}^\dagger$.
Note that, whereas the monomeric instantaneous response $\mu_{0}(t)$ is kept intact, the others are replaced by $\mu_{c.m.}(t)\simeq 2\delta (t)/(N\gamma_1)$, and the internal relaxation\footnote{
Power laws obtained by the superposition of normal modes exploit the integral formula:
\begin{eqnarray}
\int_0^\infty dx\, x^{b-1}e^{-ax^\theta}
=
\Gamma(b/\theta)/(\theta a^{b/\theta}),
\label{Integral_formula}
\end{eqnarray}
where $a,\,b\,,\theta>0$.
Note that a conventional notation $\Gamma(\cdot)$ for the gamma function is employed in eq.~(\ref{Integral_formula}), which is the same as the frictional kernel in eq.~(\ref{GLE_p}); however, it carries no special significance.
}
\begin{eqnarray}
\mu_\alpha(t)
&=&
- \sum_{q\geq 1} \frac{k_q}{\gamma_q^2} e^{-(k_q/\gamma_q)t} (h_{q,n}^\dagger)^2
\simeq 
-\frac{1}{\tau_u\gamma} \left| \frac{t}{\tau_u} \right|^{-2+(2/z)}.
\label{mu_viscous}
\end{eqnarray}
When calculating the MSD with the internal relaxation kernel eq.~(\ref{mu_viscous}), we encounter the subdiffusion $\left< \Delta x(t)^2 \right>\sim t^{2/z}$, which is a general expression.
In fact, for the dynamical exponent $z=4$ in the Rouse model, we rediscover the subdiffusion $\left< \Delta x(t)^2 \right>\sim t^{1/2}$.

Although the polymeric parameters differ from those of the Rouse polymer, the analogous formalities to obtain the static GLE-like expression are available; we therefore arrive at the same consequence as eq.~(\ref{GLE_xn}),\,(\ref{static_FDR_u}).
Note that, as expected, the dynamical effects of the HIs eventually become irrelevant to the explicit static expression of eqs.~(\ref{GLE_xn}),\,(\ref{mu_st}).
However, the SA polymer has the long-range SA interaction, where the summation is qualitatively separated into a few cases; the consequences are as follows (see appendix for details):
\begin{eqnarray}
\left< \overline{\eta}_n\overline{\eta}_{n'} \right> 
&\simeq &
\frac{k_BT}{k} 
\sum_{q\geq 1} \frac{1}{N}
\left( \frac{\pi q}{N} \right)^{1-2\nu}
\nonumber \\
&&\times \left( \cos{\left( \frac{\pi q(n-n')}{N} \right)}-\cos{\left( \frac{\pi q(n+n')}{N} \right)} \right)
\nonumber \\
&\simeq&
\left\{
\begin{array}{ll}
\frac{k_BT}{k}  & n\simeq n' 
\\
\frac{k_BT}{k} |n-n'|^{2\nu-2} & |n-n'|\gg 1
\end{array}
\right..
\label{noise_xn_corr_specific}
\end{eqnarray}
A strategy similar to that used for the Rouse polymer can be used to estimate the stretching rate for the SA polymer; however, the physical manners should be qualitatively distinct.
Unlike the dynamical kernel $\mu(t-s)$, the translational symmetry for the static kernel $\mu_{st}(n,n')$ does not always hold in the presence of a long-range interaction for the finite chain length $N$, where effects of the chain ends can enter into eqs.~(\ref{noise_xn_corr}),\,(\ref{mu_st}).

\begin{figure}[t]
\begin{center}
\includegraphics[scale=0.5]{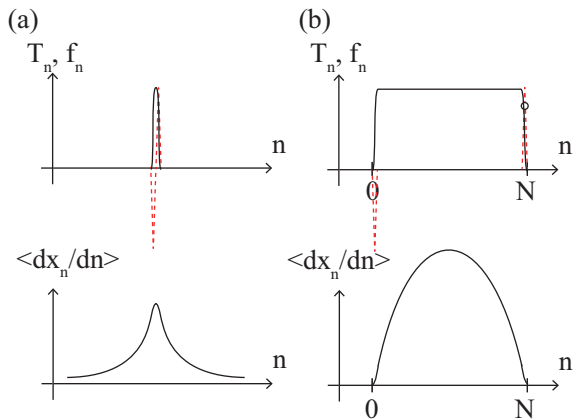}
      \caption{
		(Color online) Schematic profiles for an SA polymer.
		Axes and legends are the same as those in fig.~\ref{sketch_fig_Rouse}.
	  }
\label{sketch_fig_SA}
\end{center}
\end{figure}

For a local dipole with $f_{n}=-f\partial \delta (n-n_d)/\partial n$, as shown in fig.~\ref{sketch_fig_SA}\,(a), the stretching rate $d\left< x_n\right>/dn \simeq (k_BTf/k) |n-n_d|^{2\nu-2}$ for $n\neq n_d$ is found to have a long tail because of the long-range SA interaction.
Because $2\nu-2=-1/2$ for two dimensions $(\nu=3/4)$, or $2\nu-2\simeq -4/5$ for three dimensions $(\nu \simeq 3/5)$, the monomers are increasingly stretched as they approach the $n_d$-th monomer, i.e., $d\left< x_n\right>/dn~\nearrow$ as $|n-n_d|\searrow$. 
In addition, as shown in fig.~\ref{sketch_fig_SA}\,(b), pulling both the ends with $T_n=f\Theta (n-\epsilon)-f\Theta (n-N+\epsilon)$ creates global inhomogeneous stretching $d\left< x_n\right>/dn \simeq (f/k) [n^{2\nu-1}+(N-n)^{2\nu-1}]$.
Given that $2\nu-1=1/2$ for two dimensions, or $\simeq 1/5$ for three dimensions, the stretching rate $d\left< x_n\right>/dn$ increases as the monomers approach the $N/2$-th monomer along the chain, i.e., $d\left< x_n\right>/dn~\nearrow$ as $|n-N/2|\searrow$.

We have thus far seen the analogous GLE-like form even in the static situations; however, a noticeable difference is observed with respect to the absence of the causality on the static formulation. In fact, the static response exhibits the approximate reversal symmetry (not the exact symmetry because of the chain end effects, unless $n=N/2$), e.g., as fig.~\ref{sketch_fig_SA}\,(a), whereas a response is temporally followed by an instantaneous stimulus in the dynamical GLE.

\section{Partition functions}
\label{micro_der}

We here address the same statistical issue to relate a static response with fluctuations as that of eq.~(\ref{static_FDR_u}) viewed from the partition function based on the path integral.
The general form of the partition function for the observation of $x_n$ is given by
\begin{eqnarray}
Z_x[\{ T_n \}]
&=& \int Dx_n\,\exp{\left( -\frac{U[\{ x_n \}] -\int_0^N dn\, T_n\partial_n x_n}{k_BT} \right)},
\nonumber \\
\label{partition_Zx}
\end{eqnarray}
where an additional term $\int_0^N dn\, T_n\partial_n x_n$ is inherent in the polymer with $\partial_n x_n$ representing the local stretching rate. 
The conjugate quantity to $x_n$ is $T_n=-\int_0^ndn'\,f_{n'}$ is defined as minus the sum of the applied force from one end with $\int_0^Ndn\,\partial_n T_n=0$ retained, and $T_n$ is interpreted as applied tension acting on the $n$-th monomer. Note that the coarse-grained monomeric description of $U(\{ x_n \})$ can be given by the Edwards Hamiltonian,\footnote{
Incorporation of the SA interaction into the bonding potential can be provided by the Edwards Hamiltonian:
\begin{eqnarray}
U(\{ x_n \})
&=&
\int_0^N dn\,
\frac{1}{2}k \left(\frac{\partial x_n}{\partial n}\right)^2
\nonumber\\
&&+
v_{SA} \int_0^Ndn \int_0^Ndn'\, \delta (x_n-x_{n'}),
\label{Edwards_Hamiltonian}
\end{eqnarray}
where $v_{SA}$ denotes the SA parameters.
}
which formally has integration over $n$; thus, a double integral $\int dn\int dn'$ is hidden in the argument of the exponential for $Z_x$ through $U(\{ x_n \})$.

The role of the additional term $-\int_0^N dn\, T_n\partial_n x_n$ is clearer if we consider the local applied dipole force $f_{n}=-f\partial \delta (n-n_d)/\partial n$ given by the tension $T_{n}=f\delta (n-n_d)$. 
This tension contributes to the Hamiltonian by {$-\int_0^N dn\, T_n\partial_n x_n=f \partial_n x_n|_{n=n_d}$, which indicates the local stretching at $n=n_d$. 
In addition, given the uniform tension under $T_n=f\Theta (n-\epsilon)-f\Theta (n-N+\epsilon)$, we characterize the typical tensile-deformed size as
\begin{eqnarray}
\left<  \Delta x  \right>_f
&=&
\int Dx_n\,\Delta x 
\frac{\exp{\left( -\frac{U[\{ x_n \}] -f\Delta x}{k_BT} \right)} }{Z_x},
\label{mean_x_BothPull}
\end{eqnarray}
where $\Delta x\equiv x_N-x_0$ simply denotes the end-to-end distance and $\langle ~\rangle_f$ represents the ensemble average at the constant applied force $f$.
As in a general equilibrium relation $k_BT^2 \partial \left< U \right>/\partial T=\left< \delta U^2 \right>$~\cite{Kubo_problem}, standard calculus with canonical ensemble leads to a static FRR:
\begin{eqnarray}
\frac{\partial \left<  \Delta x\right>_f}{\partial f}
&=&
\frac{\left< \delta \Delta x^2\right>_f }{k_BT}.
\label{FRR_staric_x}
\end{eqnarray}
Note that eq.~(\ref{FRR_staric_x}) or $k_BT^2 \partial \left< U \right>/\partial T=\left< \delta U^2 \right>$ represents a response to mechanical or thermal perturbation, respectively.
Incidentally, when $fx/k_BT\ll 1$, eq.~(\ref{FRR_staric_x}) is linearized to $\left< \Delta x \right>_f/f\simeq \left< \Delta x^2 \right>_0/k_BT$, which is considered a form more analogous than eq.~(\ref{FRR_staric_x}) to the dynamical FRR $\left< \Delta x(t) \right>/f=\left< \delta \Delta x(t)^2 \right>/k_BT$ satisfied by the displacement~\cite{PRE_Sakaue_2013,PRE_Saito_2015}.

We next verify eq.~(\ref{static_FDR_u}).
Consider the thermodynamic function ${\cal F}_x\equiv -k_BT\log{Z_x}$.
The functional derivative of ${\cal F}_x$ with respect to $T_n$ results in
\begin{eqnarray}
\left<  \partial_n x_n  \right>_f
&=&
\frac{1}{Z_x}
\int Dx_n\, \partial_n x_n 
\nonumber \\
&&
\times 
\exp{\left( -\frac{U(\{ x_n \}) -\int_0^N dn\, T_n\partial_n x_n}{k_BT} \right)}.
\label{functional_dx_mean}
\end{eqnarray}
Again, applying $\delta/\delta T_m$ to eq.~(\ref{functional_dx_mean}), we arrive at
\begin{eqnarray}
\frac{\delta \left<  \partial_n x_n  \right>_f}{\delta T_m}
&=&
\frac{\left< (\partial_nx_n-\left< \partial_nx_n\right>) (\partial_mx_m-\left< \partial_mx_m\right>) \right>}{k_BT}
\label{functional_dx_corr}
\end{eqnarray}
Recall that the static kernel appearing in the GLE (eq.~(\ref{GLE_xn})) is defined as a functional derivative of $\left<\partial_n x_n\right>$ with respect to $f_m=-\partial_m T_m$ (i.e., $\mu_{st}(n-m)=\delta \left< \partial_n x_n \right>_f/\delta f_m=-\delta \left< \partial_n x_n \right>_f/\delta (\partial_m T_m)$).
A functional derivative of the mean trajectory for the GLE~(\ref{GLE_xn}) given by
$\partial_n \left< x_n\right> = -\int_0^N dn'\, \mu_{st}(n,n') \partial_{n'} T_{n'}$ is transformed through integration by parts,\footnote{
A functional derivative with respect to $T_m$ leads to
\begin{eqnarray}
\frac{\delta (\partial_n \left< x_n\right>)}{\delta T_m} 
=
- \int_0^N dn'\, \mu_{st}(n,n') 
\frac{\delta (\partial_{n'} T_{n'}) }{\delta T_m}
\end{eqnarray}
Under integration by parts, it is rewritten as
\begin{eqnarray}
\frac{\delta (\partial_n \left< x_n\right>)}{\delta T_m} 
&=&
\left[
-\mu_{st}(n,n') 
\frac{\delta T_{n'}}{\delta T_m}
\right]_{n'=0}^{n'=N}
\nonumber \\
&&
+\int_0^N dn'\, \partial_{n'} \mu_{st}(n,n') 
\frac{\delta T_{n'}}{\delta T_m}
\nonumber \\
&=&
\partial_{m} \mu_{st}(n,m),
\end{eqnarray}
where the fixed end condition $T_0=T_N=0$ is used.
}
resulting in $\delta \left< \partial_n x_n \right>_f/\delta T_m = \partial_m \mu_{st}(n,m)$.
Substituting it into eq.~(\ref{functional_dx_corr}) gives
\begin{eqnarray}
k_BT\partial_m \mu_{st}(n,m)
&=&
\left< (\partial_nx_n-\left< \partial_nx_n\right>) (\partial_mx_m-\left< \partial_mx_m\right>) \right>
\nonumber \\
\label{FRR_partition_function}
\end{eqnarray}
The resultant equation~(\ref{FRR_partition_function}) is the same as FRR eq.~(\ref{static_FDR_u}).

We emphasize that the above derivation of eq.~(\ref{static_FDR_u}) does not use the mode analyses; thus, the above argument does not rely on the assumption imposed on the mode analyses.

\section{Stochastic energetics analogy}
\label{SE_analogy}

The notion of heat based on the Langevin dynamics proposed by Sekimoto~\cite{SekimotoBook} has critically contributed to the development of stochastic energetics or stochastic thermodynamics.
Here, a question arises: What analogous relations do we obtain by replacing the $t$-axis with the $n$-axis in these frameworks?
This section is devoted to an analogy of the first law of thermodynamics.

In discussing the energetics analogy, the form of the static noise correlation (eq.~(\ref{static_FDR_u})) warrants attention.
The static noise correlation $\left< \overline{\eta}_n \overline{\eta}_{m} \right> =k_BT \partial_{m} \mu_{st}(n,m)$ (eq.~(\ref{static_FDR_u})) is no longer directly proportional to the response function $\mu_{st}(n,m)=\delta \left< dx_{n}/dn \right>/\delta f_{m}$ like the dynamical FRR $\left< \eta(t)\eta(s) \right>=k_BT\mu(t-s)$ (eq.~(\ref{GLE_x_FDR})) with $\mu(t-s)=\delta \left< dx(t)/dt \right>/\delta f(s)$; it is instead proportional to its derivative $\partial_{m} \mu_{st}(n,m)$. In addition, the symmetricity between $n$ and $m$ does not generally hold (i.e., $\mu_{st}(n,m)\neq \mu_{st}(m,n)$) and the causality is absent. 
Thus, the formalism of the noise characteristics is qualitatively distinct and some modifications are therefore required.

Sekimoto's heat definition is consistent with that in the fluctuation theorem~\cite{PRE_Crooks_1999,RPP_Seifert_2012}.
Our present strategy is to find the analogous Sekimoto heat that is compatible with the mathematical expression appearing in the fluctuation theorem. 

We begin by attempting to find the common mathematical form shared with the fluctuation theorem.
When a particle undergoes a stochastic motion in the presence of thermal agitation, the fluctuation theorem associates heat with a logarithmic function of the ratio between the probability of a forward path and that of a reverse path.
We consider, analogously, a forward path of the polymer configuration $x_0 \rightarrow x_1 \rightarrow \cdots \rightarrow x_N$ and the reverse path $x_N \rightarrow \cdots \rightarrow x_1 \rightarrow x_0$.
We then introduce a logarithmic function of the ratio between the configurational path probabilities:
\begin{eqnarray}
B(N)
\equiv 
\log{\frac{{\cal P}[x_{(\cdot)}|x_0]}{{\cal P}[x_{(\cdot)}^\dagger|x_0^\dagger]}},
\label{Q_analogue}
\end{eqnarray}
where ${\cal P}[x_{(\cdot)}|x_0]$ represents the conditional probability of the forward path $\{x_{(\cdot)}\}=\{x_1 \rightarrow x_2 \rightarrow \cdots \rightarrow x_{N-1} \rightarrow x_N\}$ given $x_0$ and ${\cal P}[x_{(\cdot)}^\dagger|x_0^\dagger]$ denotes the conditional probability of the reverse path $\{x_{(\cdot)}^\dagger\}=\{x_{N-1} \rightarrow x_{N-2} \rightarrow \cdots \rightarrow x_1 \rightarrow x_0\}$ given $x_0^\dagger \equiv x_N$.

The next step is to incorporate the Onsager--Machlup approach into eq.~(\ref{Q_analogue}).
The probability of the forward path given one end position $x_0$ is written with a sequence of noise $\{\overline{\eta}_{(\cdot)}\}=\{\overline{\eta}_1,\overline{\eta}_2,\cdots,\overline{\eta}_N\}$:
\begin{eqnarray}
{\cal P}[\overline{\eta}_{(\cdot)}]
&\sim&
\exp{\left(-\int_0^N dn\int_0^N dn'\,\frac{ (\partial_{n'}\mu_{st})^{-1}(n,n')}{2k_BT}\overline{\eta}_n\overline{\eta}_{n'} \right)},
\nonumber \\
\label{Pr_eta}
\end{eqnarray}
where $(\partial_{n'}\mu_{st})^{-1}(n,n')$ denotes the Green's function of $\partial_{n'}\mu_{st}(n,n')=\partial_{n}\mu_{st}(n',n)$ appearing on the right-hand side of eq.~(\ref{static_FDR_u}), preserving the symmetricity $(\partial_{n'}\mu_{st})^{-1}(n,n')=(\partial_{n}\mu_{st})^{-1}(n',n)$.
Similarly, we have the probability of the reverse sequence ${\cal P}[\overline{\eta}_{(\cdot)}^{(R)}|x_0^\dagger]$, where the sequence of noise $\overline{\eta}_{(\cdot)}^{(R)}$ is generated so that we can trace the reverse sequence $x_{N-1} \rightarrow x_{N-2} \rightarrow \cdots \rightarrow x_0$.
The sequence of noise in the reverse path is represented by the sequence of the forward path (see appendix). 
Keeping this in mind, we then make a logarithmic function of the ratio of these probability distributions, whose variables are converted from the sequence of noise $\{ \overline{\eta}_{(\cdot)} \}$ to $\{x_{(\cdot)}\}$ (or $\{ \overline{\eta}_{(\cdot)}^{(R)} \}$ to $\{x_{(\cdot)}^\dagger\}$); the logarithmic function is found to be\begin{eqnarray}
\log{
\frac{{\cal P}[x_{(\cdot)}|x_0]}{{\cal P}[x_{(\cdot)}^\dagger|x_0^\dagger]}
}
&=&
2\int_0^N dn\int_0^N dn'\, \frac{(\partial_{n'}\mu_{st})^{-1}(n,n')}{k_BT}
\nonumber \\
&& \times \left( \frac{dx_{n'}}{dn'} -\overline{\eta}_{n'}\right)\frac{dx_n}{dn}.
\label{ratio_P_n}
\end{eqnarray}
The factor ``2" is included in front of the integrals because of the absence of the causality observed in the temporal evolution.
The right-hand side appears similar to the heat in Sekimoto's definition, and we arrive at a possible candidate for the analogous heat $\overline{Q}=k_BTB(N)/2$.

To proceed further, we need a technical preparation for the inverse kernel $(\partial_{n'}\mu_{st})^{-1}(n,n')$ appearing in eq.~(\ref{ratio_P_n}).
This kernel is associated with the inverse kernel of $\mu_{st}(n,m)$ (i.e., $\Gamma_{st} (n,m) \equiv \mu_{st}^{-1}(n,m)$).
Applying $\Gamma_{st} (n,m)$ to eq.~(\ref{GLE_xn}), we obtain a static GLE expression paired to eq.~(\ref{GLE_xn}) as 
\begin{eqnarray}
f_n
=\int_0^N dn'\, \Gamma_{st}(n,n')\frac{dx_{n'}}{dn'} +\overline{\eta}_n^{(f)}.
\label{GLE_fn}
\end{eqnarray}
Recall that the static kernels $\Gamma_{st} (n,m)$ are defined such that $\int_0^Ndm\,\Gamma_{st} (n,m)\mu_{st}(m,n')=\delta(n-n')$ and such that a simple calculation verifies $\overline{\eta}_n^{(f)}=-\int_0^Ndm\, \Gamma_{st}(n,m)\overline{\eta}_{m}$ (see appendix).
In addition, differentiating the definitional identity $\int_0^Ndm\,\Gamma_{st} (n,m)\mu_{st}(m,n')=\delta(n-n')$ with respect to $n'$ and integrating it with respect to $n$, we obtain
$\int_0^Ndm\,\left[\int_0^n dm'\,\Gamma_{st} (m',m)\right]\partial_{n'} \mu_{st}(m,n')=\delta(n-n')$.
A comparison of this expression with the definition then implies\footnote{
The explicit expressions in the mode space are written as
\begin{eqnarray}
\partial_{n'} \mu_{st}(n,n') 
&=& \sum_{q \geq 1} \frac{1}{k_{q}} \left( \frac{\pi q}{N} \right)^2 h_{q,n'}^{(s)}h_{q,n}^{(s)^\dagger}
\\
\int_0^n dm\,
\Gamma_{st}(m,n)
&=&
\sum_{q \geq 1} k_{q} \left( \frac{\pi q}{N} \right)^{-2} h_{q,n}^{(s)}h_{q,m}^{(s)\dagger}.
\end{eqnarray}
}
\begin{eqnarray}
[\partial_{n'} \mu_{st}]^{-1}(n,n') 
=
\int_0^ndm'\,\Gamma_{st}(m',n').
\label{inverse_del_mu}
\end{eqnarray}
To utilize this expression, we rephrase the static noise correlation of eq.~(\ref{static_FDR_f}) (see appendix) with $-\int_0^ndm\,\Gamma_{st}(m,n')$ and $\overline{\eta}_m^{(f)}= -\int_0^Ndm'\, \Gamma_{st}(m,m')\overline{\eta}_{m'}$ as
\begin{eqnarray}
&&\left< \left[\int_0^ndm\,\overline{\eta}_m^{(f)}\right]\left[\int_0^{n'}dm'\,\overline{\eta}_{m'}^{(f)}\right] \right>
\nonumber \\
&=&
-k_BT\int_0^ndm''\, \Gamma_{st}(m'',n').
\label{static_FDR2nd}
\end{eqnarray}
Equation~(\ref{static_FDR2nd}) is the symmetric form pertinent to the FRR (eq.~(\ref{GLE_x_FDR})).
Indeed, $\int_0^ndm\, \Gamma_{st}(m,n')$ is interchangeable between $n$ and $n'$.
Note that, from eqs.~(\ref{static_FDR_u}),\,(\ref{Rouse_static_noise_corr}),\,(\ref{inverse_del_mu}), we find that $\int_0^ndm\,\overline{\eta}_m^{(f)}$ provides the stationary noise in the Rouse polymer along the $n$-axis.

We now progress to the final step in the energetics analogy from the viewpoint of force balance.
Recalling the force balance of eq.~(\ref{GLE_fn}), we integrate it with respect to $n$ and organize it as
\begin{eqnarray}
T_n
&=&
\int_0^N dn'\, \left[ -\int_0^ndm\, \Gamma_{st}(m,n')\right]\frac{dx_{n'}}{dn'} -\int_0^ndn'\,\overline{\eta}_{n'}^{(f)},
\nonumber \\
\label{modified_FB}
\end{eqnarray}
where $T_n=-\int_0^ndn'\,f_{n'}$ is used.
The right-hand side of eq.~(\ref{modified_FB}) is then found to be hidden in eq.~(eq.~(\ref{ratio_P_n})) because 
\begin{eqnarray}
\log{
\frac{{\cal P}[x_{(\cdot)}|x_0]}{{\cal P}[x_{(\cdot)^\dagger}|x_0^\dagger]}
}
&=&
\frac{2}{k_BT}\int_0^N dn
\nonumber \\
&& \times
\int_0^N dn'\,\left[ \int_0^ndm'\,\Gamma_{st}(m',n')\right] \frac{dx_{n'}}{dn'} \frac{dx_n}{dn}
\nonumber \\
&&
+
\frac{2}{k_BT}\int_0^N dn\, \left[ \int_0^{n}dm'\, \overline{\eta}_{m'}^{(f)}\right]\frac{dx_n}{dn}
\label{ratio_P_n_Gamma}
\end{eqnarray}
with  $\int_0^ndm'\,\overline{\eta}_{m'}^{(f)}=-\int_0^ndm'\int_0^Ndm''\, \Gamma_{st}(m',m'')\overline{\eta}_{m''}=-\int_0^Ndm''\, [\partial_{m''}\mu_{st}]^{-1}(n,m'')\overline{\eta}_{m''}$ and eq.~(\ref{inverse_del_mu}).
We are naturally led to the definition of the analogous heat as
\begin{eqnarray}
d'\overline{Q}_n
&\equiv&
\Biggl( \int_0^Ndn'\,\left[\int_0^ndm'\, \Gamma_{st}(m',n')\right]\frac{dx_{n'}}{dn'}
\nonumber \\
&&+\int_0^ndn'\,\overline{\eta}_{n'}^{(f)} \Biggr) dx_n.
\label{analogue_heat}
\end{eqnarray}
The kernel part in the first term and the integrated noise in the brackets on the right-hand side satisfy the symmetric analogous FRR (eq.~(\ref{static_FDR2nd})).
Equation~(\ref{analogue_heat}) is an analogous form of heat of a Brownian particle defined by Sekimoto~\cite{SekimotoBook}, and, to be precise, eq.~(\ref{analogue_heat}) is defined like a non-Markov process, as in refs.~\cite{EPL_Harada_2005,JStatMech_Ohkuma_Ohta_2007}, with the GLE. In addition, as in the definition by Sekimoto, the Stratonovich multiplication is implicitly employed in eq.~(\ref{analogue_heat}) although the product notation is not explicit.
Furthermore, the integration $\overline{Q}\equiv \int_{n=0}^{n=N} d'\overline{Q}_n$ yields $\overline{Q}=k_BTB(N)/2$ expected before; then, the definition of $B(N)$ and the polymer configuration identity\footnote{The joint probability density ${\cal P}(x_0,x_1,\cdots,x_N)$ indicates
\begin{eqnarray}
{\cal P}(x_0,x_1,\cdots,x_N)
=
{\cal P}[x_{(\cdot)}|x_0]{\cal P}(x_0)
=
{\cal P}[x_{(\cdot)}^\dagger|x_0^\dagger]{\cal P}(x_{N}),
\label{DB_stretching}
\end{eqnarray}
where ${\cal P}(x_i)$ denotes the probability density for $x_i$ for $i=0$ or $N$ and $\dagger$ represents the reverse configuration path with $x_n^\dagger=x_{N-n}$ and $f_n^\dagger=-f_{N-n}$. For simplicity, the applied force $f_n$ or $f_n^\dagger$ is not shown in the arguments (see also appendix).
One might think nonequilibrium conditions no longer validate ${\cal P}[x_{(\cdot)}|x_0]{\cal P}(x_0)={\cal P}[x_{(\cdot)}^\dagger|x_0^\dagger]{\cal P}(x_{N})$; however, it always holds because of the characteristics of the one-dimensional chain structure.
Equation~(\ref{DB_stretching}) multiplied by the Boltzmann constant $k_B$ reads
\begin{eqnarray}
k_B B(N)
=
-k_B\log{{\cal P}(x_0)}
-(-k_B\log{{\cal P}(x_N)}),
\end{eqnarray}
which shows the difference in the Shannon entropy of the chain ends.
}
suggest $T\Delta S=-2\overline{Q}$ associated with a difference in Shannon entropy between the end monomers $\Delta S$.

A caveat is symmetricity between the forward and reverse paths.
Recall the dynamical Langevin equation for $t\in [0,T]$ during time interval $T$, where the context of the fluctuation theorem supposes $x^\dagger(t)=x(T-t)$ and $dx^\dagger(t)/dt=-dx(t')/dt'|_{t'=T-t}$ with the analogous notation of the reverse path defined like those around eq.~(\ref{Q_analogue}).
Similarly, the forward and the reverse paths along the polymer configuration are associated through $x^\dagger_n=x_{N-n}$ and $dx^\dagger_n/dn=-dx_{n'}/dn'|_{n'=N-n}$.
However, inspecting the tension reveals the manifestation of the distinct symmetricity inherent to the static Langevin equation, where the tension is not altered under the labeling conversion (i.e., $T^\dagger_n=T_{N-n}$). 
This result indicates odd symmetricity of the applied force $f^\dagger_n=-f_{N-n}$ because $f_n=-dT_n/dn$.
If the reverse paths were considered in figs.~\ref{sketch_fig_Rouse},\,\ref{sketch_fig_SA}, $f_n^\dagger=-f_{N-n}$ would be introduced to keep the tension $T^\dagger_n=T_{N-n}$ unchanged.
Also, $q$-dependent symmetricity is hidden in the kernel $\Gamma_{st}(n,n')$ (see eq.~(\ref{Gamma_st}) in appendix).
Thus, the different symmetry among $f_n$ and $\int_0^Ndn'\,\Gamma(n,n')dx_{n'}/dn'$ coexists in eq.~(\ref{GLE_fn}) under the labeling conversion $n\leftrightarrow N-n$, which implies that sequences of forward and reverse noise are not generally identical (i.e., $\overline{\eta}_n^{(f)\dagger}\neq \overline{\eta}_{N-n}^{(f)}$).

A remaining issue is how to interpret $T_n\partial_nx_n=T_n(dx_n/dn)$ under the first law of thermodynamics.
Two definitions are discussed here.
The first definition considers a source of $T_n\partial_nx_n$ as an external origin, and the analogous infinitesimal work is defined as
\begin{eqnarray}
d'\overline{W}_n \equiv T_n \frac{d x_n}{d n}.
\label{def_analogy_W}
\end{eqnarray}
The total work is also obtained by $\overline{W}\equiv \int_{n=0}^{n=N} d'\overline{W}_n$.
Note that the analogous work done by the 
exterior 
is assigned to be positive because the direction of the conventional work done is arbitrary. 
We then arrive at the analogue of the energy balance equation:
\begin{eqnarray}
\overline{W}-\overline{Q}=0.
\label{analogy_EB_W_Q}
\end{eqnarray}
The internal energy in eq.~(\ref{analogy_EB_W_Q}) appears absent although Langevin eq.~(\ref{x_mode}) at the outset explicitly includes the conservative force produced by the potential $U[\{ x_n \}]$.
The effective elasticity created by the polymer chain is, however, implicitly embedded into the static kernel $\Gamma_{st}(n,n')$ in the analogous generalized Langevin eq.~(\ref{GLE_fn}).

Under the other definition, $T_n\partial_nx_n$ is considered part of the interior.
This picture defines the internal energy:
\begin{eqnarray}
U_T \equiv -\int_0^N dn\, T_n\frac{dx_n}{dn},
\end{eqnarray}
which appears as the second term in $U[\{ x_n \}] -\int_0^N dn\, T_n\partial_n x_n$ of exponential eq.~(\ref{partition_Zx}).
We then arrive at the analogue of the energy balance equation: 
\begin{eqnarray}
\Delta U_T = -\overline{Q},
\label{analogy_EB_U_Q}
\end{eqnarray}
where $\Delta U_T\equiv U_T-U_T|_{\{\partial_1 x_1=\partial_2 x_2=\cdots=\partial_N x_N=0\}}$.
Similarly to the first definition, the conservative force produced by the potential $U[\{ x_n \}]$ is incorporated into the static kernel $\Gamma_{st}(n,n')$, which results in the analogous heat.

\section{Nonequilibrium-steady-state analogy}
\label{NSS}

A last main issue addressed in this article is to find the analogue to the steady-state fluctuation theorem~\cite{RPP_Seifert_2012,JPhysA_Kurchan_1998,JStatPhys_Lebowitz_Spohn_1999,JCP_Gaspard_2004}.
The preceding sections, as illustrated in figs.~\ref{sketch_fig_Rouse}\,(b), \ref{sketch_fig_SA}\,(b), have thus far implicitly considered that a near-equilibrium shape is obtained by pulling both chain ends with $T_n=f\Theta(n-\epsilon)-f\Theta(n-N+\epsilon)$. However, we now apply sufficient force to induce a large deformation expressed by a sequence of tensile blobs~\cite{deGennesBook,Macromolecules_Pincus_1976,JCP_Baba_Murayama_2012}, where the interaction range is finite even in the presence of the SA effects.
This approach enables us to proceed toward the steady-state-fluctuation analogy.
However, we need to grasp the physical differences.
The monomer indices $n$ run along the chain backbone, which is largely deformed under ``equilibrium" although the steady-state fluctuation theorem discusses the temporal evolution of the system under a ``nonequilibrium" steady state.
Nonetheless, a basic mathematical construction is shared by simply replacing time $t$ appearing in the context of the steady-state fluctuation theorem by a chain length $N$ in the polymer.

Recall the analogous energy balance equation $\int_0^N dn\,T_{n}\partial_n x_{n}=\overline{Q}$.
The tension serves as the driving force to maintain the nonequilibrium steady state in the conventional steady-state fluctuation theorem.

The following analytical procedure mainly refers to the literature related to the steady-state fluctuation theorem~\cite{JStatPhys_Lebowitz_Spohn_1999,JCP_Gaspard_2004}, although other important works are not cited here.
Using the fluctuating quantity $B(N)$, we consider a generating function with parameter $\lambda$:\footnote{
Note that the probability density for the end configuration $x_0$ is assumed to be irrelevant to the resultant divided by $N$ for sufficiently large $N$ such as
\begin{eqnarray}
&& \frac{1}{N}\log{\left< e^{-\lambda B(N)} \right>}
\nonumber \\
&\equiv&
\frac{1}{N}\log{
\left[
\int Dx_{(\cdot)} 
\int dx_0\, 
{\cal P}[x_{(\cdot)}|x_0] {\cal P}(x_0) e^{-\lambda B(N)}
\right]
}
\nonumber \\
&\rightarrow&
\frac{1}{N}\log{
\left[
\int Dx_{(\cdot)}\, 
{\cal P}[x_{(\cdot)}|x_0] e^{-\lambda B(N)}
\right]
}
\nonumber \\
&=&
\frac{1}{N}\log{
\left[
\left< e^{-\lambda B(N)} \right>_{lc}
\right]
}.
\end{eqnarray}
The integrating variables in $\int Dx_{(\cdot)}\, {\cal P}[x_{(\cdot)}|x_0]e^{-\lambda B(N)}$ are tentatively converted from position $\{ x_n \}$ to the displacement $\{ \Delta x_n \}$ in the second line so that the integrations with respect to $x_0$ and $\int {\cal D}x_{(\cdot)}$ can be separated.
In the third line, the limit of $N\rightarrow +\infty$ eliminates $N^{-1}\log{\int dx_0\, {\cal P}(x_0)}$ and we again recover the variables from the displacement to the positions.}
\begin{eqnarray}
C(\lambda)
=
-\lim_{N\rightarrow +\infty} \frac{1}{N}\log{\left< e^{-\lambda B(N)} \right>_{lc}},
\label{def_C_lambda}
\end{eqnarray}
where $\left< (\cdot) \right>_{lc}$ is defined as the average taken over the chain configuration paths $\{ x_{(\cdot)} \}$ for a sufficiently ``long chain" with
\begin{eqnarray}
\left< e^{-\lambda B(N)} \right>_{lc}
\equiv
\int Dx_{(\cdot)}\, 
{\cal P}[x_{(\cdot)}|x_0] e^{-\lambda B(N)}.
\label{exp_BN}
\end{eqnarray}
Notably, eq.~(\ref{def_C_lambda}) considers the average of the negative logarithmic function of eq.~(\ref{exp_BN}) over $N$, whereas the conventional nonequilibrium steady state takes the temporal average over time period $T$. 
Symmetricity embedded in the generating function is found as
\begin{eqnarray}
\left< e^{-\lambda B(N)} \right>_{lc}
&=&
\int Dx_{(\cdot)}\, 
{\cal P}[x_{(\cdot)}|x_0]
\left(
\frac{{\cal P}[x_{(\cdot)}|x_0]}{{\cal P}[x_{(\cdot)}^\dagger|x_0^\dagger]}
\right)^{\lambda}
\nonumber \\
&=&
\int Dx_{(\cdot)}\, 
{\cal P}[x_{(\cdot)}^\dagger|x_0^\dagger]
\left(
\frac{{\cal P}[x_{(\cdot)}^\dagger|x_0^\dagger]}{{\cal P}[x_{(\cdot)}|x_0]}
\right)^{1-\lambda}
\nonumber \\
&=&
\int Dx_{(\cdot)}^\dagger\, 
{\cal P}[x_{(\cdot)}^\dagger|x_0^\dagger]
\left(
\frac{{\cal P}[x_{(\cdot)}^\dagger|x_0^\dagger]}{{\cal P}[x_{(\cdot)}|x_0]}
\right)^{1-\lambda}
\nonumber \\
&=&
\left< e^{-(1-\lambda) B(N)} \right>_{lc},
\label{dev_FT_symmetry}
\end{eqnarray}
which leads to the symmetricity for a long chain:
\begin{eqnarray}
C(\lambda)=C(1-\lambda).
\label{FT_symmetry}
\end{eqnarray}
The first derivative of the generating function gives the first moment, and then eq.~(\ref{FT_symmetry}) yields
\begin{eqnarray}
\frac{dC}{d\lambda}\Biggr|_{\lambda=0}
=
-
\frac{dC}{d\lambda}\Biggr|_{\lambda=1}
=
\lim_{N\rightarrow +\infty} \frac{1}{N} \left< B(N) \right>_{lc}.
\label{1st_moments}
\end{eqnarray}
Equation~(\ref{1st_moments}) is analogous to the mean heat rate balanced with the energy input rate in the conventional steady-state fluctuation theorem.
From $\overline{Q}=k_BTB(N)/2$ and eqs.~(\ref{modified_FB}),\,(\ref{analogue_heat}), the analogous first moment of eq.~(\ref{dev_FT_symmetry}) yields
\begin{eqnarray}
\lim_{N\rightarrow+\infty}\frac{k_BT}{2N}\left< B(N) \right>
&=&
\lim_{N\rightarrow+\infty}
\frac{1}{N}\left<\overline{Q}\right>
\nonumber \\
&=&
\lim_{N\rightarrow+\infty}
\frac{1}{N}\int_0^N dn\,\left<T_{n}\frac{dx_{n}}{dn}\right>
\nonumber \\
&=&
\lim_{N\rightarrow+\infty}
\frac{f}{N}(\left<x_N\right>-\left<x_0\right>),
\label{ana_mean_heat}
\end{eqnarray}
where $\int_0^N dn\,\left<T_{n}dx_{n}/dn\right>$ appearing in the second line is considered to be that originating from the external source from the viewpoint of the analogy.
We also arrive at this consequence (eq.~(\ref{ana_mean_heat})) by recalling the additional factor $\int_0^Ndn\,T_n\partial x_N$ in the argument of the partition function (eq.~(\ref{partition_Zx})) or $f\Delta x$ appearing in eq.~(\ref{mean_x_BothPull}).\footnote{
This note directly finds $k_BT\left< B(N)\right>=f[\left< x_N\right>-\left< x_0\right>]$ without passing through the mean analogous heat $\left<\overline{Q}\right>$, whereas $\left<\overline{Q}\right>$ appears in the first line of eq.~(\ref{ana_mean_heat}).
Let ${\cal P}(x_0,x_N)$ denote the joint probability density for $x_0$ and $x_N$.
Using the probability density ${\cal P}_\Delta(\Delta x)$ of $\Delta x=x_N-x_0$, we rewrite it in two ways from the symmetricity between $x_0$ and $x_N$ as
${\cal P}(x_0,x_N)={\cal P}(x_0){\cal P}_\Delta(\Delta x)={\cal P}(x_N){\cal P}_\Delta(-\Delta x)$,
and then the probability density for $x_0$ or $x_N$ is ${\cal P}(x_0)={\cal P}(x_0,x_N)/{\cal P}_\Delta(\Delta x)$ or ${\cal P}(x_N)={\cal P}(x_0,x_N)/{\cal P}_\Delta(-\Delta x)$, respectively.
Substituting these probability densities into $B(N)$ leads to
\begin{eqnarray}
B(N)&=&\log{{\cal P}(x_N)}-\log{{\cal P}(x_0)}
\nonumber \\
&=&\log{{\cal P}_\Delta(\Delta x)}-\log{{\cal P}_\Delta(-\Delta x)}.
\label{B_Dx}
\end{eqnarray}
Recalling eqs.~(\ref{partition_Zx}),\,(\ref{mean_x_BothPull}), we discover that ${\cal P}_\Delta(\Delta x)\sim e^{T\Delta x}= e^{f\Delta x}$ with $T$ being the applied tension. Similarly, the reverse configuration path has ${\cal P}_\Delta(-\Delta x)\sim e^{T(-\Delta x)}=e^{-f\Delta x}$.
Substituting these expressions into eq.~(\ref{B_Dx}), we obtain
\begin{eqnarray}
B(N)=2f\Delta x.
\end{eqnarray}
This average over infinitely large $N$ becomes eq.~(\ref{ana_mean_heat}).
}
The positive extension $\left<x_N\right>-\left<x_0\right>>0$ for $f>0$ implies that $\lim_{N\rightarrow +\infty} N^{-1} \left< B(N) \right>_{lc} >0$; thus, $C(\lambda)$ is found to be an upwardly convex function from eqs.~(\ref{FT_symmetry}),\,(\ref{1st_moments}).

In a subsequent step, we introduce the large-deviation function:
\begin{eqnarray}
C_*(\theta)
=
-\lim_{N\rightarrow +\infty} \frac{1}{N}\log{ \left( {\cal P}_{lc}\left[ \frac{B(N)}{N} \in (\theta,\theta+d\theta) \right]d\theta \right)},
\nonumber \\
\label{LogPro}
\end{eqnarray}
where the probability density that the fluctuating quantity $B(N)/N$ is between $\theta$ and $\theta+d\theta$ with an infinitesimal interval $d\theta$ for a long chain $N\gg1$ is denoted by ${\cal P}_{lc}\left[ B(N)/N \in (\theta,\theta+d\theta) \right]$.
For a very long chain $N \gg 1$, we write ${\cal P}_{lc}$, as in the expression
\begin{eqnarray}
{\cal P}_{lc}\left[ \frac{B(N)}{N} \in (\theta,\theta+d\theta) \right]
=
A(\theta,N) e^{-C_*(\theta)N}d\theta,
\label{LogPro_2}
\end{eqnarray}
where $A(\theta,N)$ is implied to be a normalization factor that satisfies $N^{-1}\log{[A(\theta,N)]}\rightarrow 0$ as $N\rightarrow +\infty$.
Using eq.~(\ref{LogPro_2}), we rewrite eq.~(\ref{exp_BN}) as
\begin{eqnarray}
\left< e^{-\lambda B(N)} \right>_{lc}
&=&
\int d\theta\, {\cal P}_{lc}\left[ \frac{B(N)}{N} \in (\theta,\theta+d\theta) \right]e^{-N\lambda \theta}
\nonumber \\
&=&
\int d\theta\,A(\theta,N) e^{-NC_*(\theta)-N\lambda \theta}.
\label{integral_theta}
\end{eqnarray}
In eq.~(\ref{integral_theta}), the maximum of the argument of the exponential function appearing in the integrand has a major contribution to the integration, which suggests that
\begin{eqnarray}
C(\lambda)=C_*[\theta(\lambda)]+\lambda\theta(\lambda),
\label{Legendre_1}
\end{eqnarray}
where $\theta=\theta(\lambda)$ is a function of $\lambda$ through 
\begin{eqnarray}
\lambda
=
-\frac{dC_*(\theta)}{d\theta}.
\end{eqnarray}
In addition, inverting $\lambda$ into $\theta$, we obtain the function
\begin{eqnarray}
C_*(\theta)=C[\lambda(\theta)] -\theta\lambda(\theta).
\label{Legendre_2}
\end{eqnarray}
which is also found from the Legendre transformation function.
Furthermore, noting the symmetricity in eq.~(\ref{FT_symmetry}), we obtain $C(\lambda)=C(1-\lambda)=C_*(-\theta)-(1-\lambda) \theta$ with $\theta(\lambda)=dC(\lambda)/d\lambda=dC(1-\lambda)/d\lambda=-dC(1-\lambda)/d(1-\lambda)=-\theta(1-\lambda)$.
Combining this equation with eq.~(\ref{Legendre_1}), we obtain
\begin{eqnarray}
C_*(\theta)-C_*(-\theta)
=
-\theta
\end{eqnarray}
From this equation, we discover the other analogous form to the fluctuation theorem, which shows asymptotic behavior for $N\rightarrow +\infty$ growing as
\begin{eqnarray}
\frac{
{\cal P}_{lc}\left[ \frac{B(N)}{N} \in (\theta,\theta+d\theta) \right]
}{
{\cal P}_{lc}\left[ \frac{B(N)}{N} \in (-\theta,-\theta+d\theta) \right]
}
\simeq 
e^{\theta N}.
\label{ratio_Prob_N}
\end{eqnarray}
Note that $\theta$ is interpreted as the reduced analogous heat $2\overline{Q}/(Nk_BT)$.
The analogous heat $\overline{Q}=k_BTB(N)/(2N)$ is equal to the local change in the thermodynamic potential (or the Landau free energy) per monomer as a result of pulling both the ends, which falls into a static description.
However, the corresponding quantity in the conventional steady-state fluctuation theorem is the rate of heat transfer into the surrounding media, which is associated with the dissipation mechanism.
Also, 
we note that eq.~(\ref{ratio_Prob_N}) for the long chain is expressed without the specific polymer structures.

\section{Discussion}
\label{discussion}

We here verify the analogy of a response function to a susceptibility in the Langevin representation.
Although there are various definitions of response functions to fit into the observables, one of the most frequently used response functions is defined as a response of the mean velocity to external force, formulated as
\begin{eqnarray}
R(t,s) \equiv \frac{\delta}{\delta f(s)}\left< \frac{dx(t)}{dt} \right>.
\label{R_f}
\end{eqnarray}
When this function is substituted into the conventional GLE (eq.~(\ref{GLE_x})), the mobility kernel is associated as $R(t,s)=\mu(t-s)\Theta(t-s)$. By contrast, we consider the static analogue to eq.~(\ref{R_f}).
The susceptibility can be defined as 
\begin{eqnarray}
\chi (n,m) \equiv \frac{\delta}{\delta f_m} \left< \frac{dx_n}{dn} \right>.
\label{suscep}
\end{eqnarray}
Notably, we find that the static kernel $\mu_{st}(n,m)$ corresponds to the above susceptibility by referring to the static GLE (eq.~(\ref{GLE_xn})) (i.e., $\chi (n,m)=\mu_{st}(n,m)$). Substituting eq.~(\ref{gen_coeff}) into eq.~(\ref{mu_st}) and focusing on the smallest wavenumber mode with $q=1$, we find the scaling $\chi=\mu_{st}\sim N^{2\nu}$.
As expected, then, a comparison of $\chi\sim N^{2\nu}$ with a conventional definition $\chi \sim N^\gamma$ leads to a known consequence in the framework of the Flory's mean field theory: \footnote{We conventionally use $\gamma$ as a universal exponent associated with the susceptibility; however, the same notation ``$\gamma$" is used for the frictional coefficient in dynamics unless the identifications between them are ambiguous.}
\begin{eqnarray}
\gamma=2\nu,
\label{gamma_2nu}
\end{eqnarray}
which coincides with the expression obtained using survival probability and reported in the literature~\cite{RivNuovoCinento_Peliti_1987}. Equation~(\ref{gamma_2nu}) may not be exact but offers a good approximation.

Another issue of the analogy is fluctuations of conjugate variables.
In the dynamical fluctuations with $\Delta p(t)$ being the time integral of the applied force at the controlled position, $\Delta p(t)$ is considered a conjugate variable to $\Delta x(t)$ in the action (dimensions of [energy]$\times$[time]$=$[displacement]$\times$[momentum transfer]) or in the GLE (eq.~(\ref{GLE_x})); it undergoes superdiffusion $\left< \Delta p(t)^2 \right>\sim t^{\alpha^{(p)}}$ with $\alpha^{(p)}>1$~\cite{PRE_Saito_Sakaue_2017,PRE_Saito_2017,Polymers_Saito_Sakaue_2019}.
The index for the MSD is here rewritten with $\alpha \rightarrow \alpha^{(x)}$ to make dual expressions appear symmetric; the exponents of subdiffusive $\alpha^{(x)}$ and superdiffusive $\alpha^{(p)}$ are then associated through 
\begin{eqnarray}
\alpha^{(x)}+\alpha^{(p)}=2.
\label{alpha_x_p}
\end{eqnarray}
However, in the static GLE-like form discussed thus far, the tension $T_n$ serves as a conjugate variable to $x_n$ in the energy (dimension of [displacement]$\times$[force]) (see appendix).
We here specifically consider a system that fixes both chain-end positions by imposing the applied force $f_n^*(t)=f_0^*(t)\delta (n-\epsilon)+f_N^*(t)\delta (n-N+\epsilon)$, where the applied tension is found to be $T_n^*(t)=-f_0^*(t)\Theta (n-\epsilon)-f_N^*(t)\Theta (n-N+\epsilon)$.
Recall that imposing the constant force $f_0=-f$ and $f_N=f$ in figs.~\ref{sketch_fig_Rouse}\,(b),\,\ref{sketch_fig_SA}\,(b) does not fix the chain-end positions but provides a way to observe the position fluctuations.
By contrast, the applied force $f_0^*(t)$,\,$f_N^*(t)$ to fix the positions temporally varies and the force fluctuation is observed.
A caveat is that the force $f_n^*(t)$ applied to fix the positions has two components: (i) the center of mass and (ii) the internal structure.
To discover a static counterpart to eq.~(\ref{alpha_x_p}), we need to subtract the center-of-mass component.
Indeed, this case does not always satisfy the balance of the applied force; i.e., generally, $T_N^*(t)=-\int_0^N dn\,f^*_n(t)=-(f_0^*(t)+f_N^*(t))\neq0$, unlike that in sec.~\ref{GLEanalogy}.
The fluctuation of the total applied force is offset by $\left< [f_N^*(t)+f_0^*(t)]^2 \right>$ so that $\left< T_{2\epsilon}(t)^2 \right> = \left< T_{2\epsilon}^*(t)^2 \right> -\left< [f_N^*(t)+f_0^*(t)]^2 \right>=\left< [-2f_0^*(t)f_N^*(t)-f_N^*(t)^2] \right>$ serves as the internal-mode fluctuations.
Note that the tension is technically monitored for $\epsilon<n<N-\epsilon$, like the tension in figs.~\ref{sketch_fig_Rouse}\,(b),\,\ref{sketch_fig_SA}\,(b), and that, for example, the near-end points at $n=2\epsilon$ or $n=N-2\epsilon$ can be chosen as representative observation points.
The fluctuations $\left< \Delta T_{2\epsilon}(t)^2 \right>=\left< \Delta T_{N-2\epsilon}(t)^2 \right>$ exhibit the power law characterized by index $\nu^{(f)}$ in an analogous form to eq.~(\ref{x_static}):
\begin{eqnarray}
\left< \Delta T_{2\epsilon}^2 \right> =\left< \Delta T_{N-2\epsilon}^2 \right>\sim N^{2\nu^{(f)}},
\label{MST_n}
\end{eqnarray}
Although the spatial fluctuations are associated with the inverse spring constants as $\left< \Delta x_N^2 \right>\simeq k_BT/(Nk_1)\sim N^{2\nu^{(x)}}$ in eq.~(\ref{x_static}) with the Flory exponent rewritten as $\nu \rightarrow \nu^{(x)}$ to make the notation symmetric, the force fluctuations in eq.~(\ref{MST_n}) are converted as $\left< \Delta T_{2\epsilon}^2 \right>=\left< \Delta T_{N-2\epsilon}(t)^2 \right> \simeq k_BTNk_1\sim N^{-2\nu^{(x)}}$.
Thus, the static indices have an analogous but not identical relation to eq.~(\ref{alpha_x_p}): \begin{eqnarray}
\nu^{(x)}+\nu^{(f)}=0,
\label{complementary_static}
\end{eqnarray}
where $\nu^{(f)}<0$ is negative, or said to be ``ultrasubdiffisive" because the exponent is less than zero, let alone less than $\nu=1/2$ for the normal diffusion. Both eqs.~(\ref{x_static}) and (\ref{MST_n}) allude to the greater susceptibility that embodies the ``soft" description of soft matter because a long chain becomes flexible with a small spring constant irrespective of constraints or boundary conditions. The difference in the sum of the indices ($=2$ or $=0$) arises from the dynamical FRRs (eq.~(\ref{GLE_x_FDR}),\,(\ref{GLE_p_FDR})) and the static FRRs (eqs.~(\ref{static_FDR_u}),\,(\ref{static_FDR_f})).

Thus far, we have mainly discussed the analogy formalism.
A last argument concerns a possible application to experiments.
An interesting candidate is a crumpled globule~\cite{EPL_Grosberg_1993} used as a chromatin model~\cite{Science_Grosberg_1993}.
Although chromatin exhibits dynamical evolution,
the static analyses can be developed if the dynamics are sufficiently slow for chromatin to be considered a thermally stable structure.
An assignment with $\nu=1/3$ in GLEs~(\ref{GLE_xn}),\,(\ref{GLE_fn}) and FRRs~(\ref{static_FDR_u}),\,(\ref{static_FDR_f}) enables us to apply the present basic formulation.  
Using bold font to represent vectors, e.g., ${\bf x}_n=(x_n,y_n,z_n)$ in a Cartesian coordinate system, we modify the static Langevin equation from one to three dimensions to represent the steric configurations:
\begin{eqnarray}
\frac{d{\bf x}_{n}}{dn}=\int_0^Ndn'\,\mu_{st}(n,n'){\bf f}_{n'}+\overline{\boldsymbol{\eta}}_{n},
\label{3D_staticGLE}
\end{eqnarray}
where the noise components are assumed to have independent Cartesian components that satisfy the FRR as $\left< \overline{\boldsymbol{\eta}}_{n} \cdot \overline{\boldsymbol{\eta}}_{n'} \right>=3k_BT\partial_{n'} \mu_{st}(n,n')$ with the numerical coefficient $3$ accounting for three dimensions.
The assumption of a Gaussian distribution in eq.~(\ref{3D_staticGLE}) provides the conditional probability density given ${\bf x}_{0}$:
\begin{eqnarray}
P({\boldsymbol x}_n|{\boldsymbol x}_0)  
&=&
\frac{1}{\sqrt{2\pi \left< \Delta \boldsymbol{x}_n^2 \right>_0 }}
\exp{} \Biggl(
\nonumber \\
&& 
-\frac{\left( \Delta {\boldsymbol x}_n-\int_0^n dm\int_0^N dn'\,\mu_{st}(m,n'){\boldsymbol f}_{n'} \right)^2}{2\left< \Delta \boldsymbol{x}_n^2 \right>_0}
\Biggr),
\nonumber \\
\label{PDF_xn}
\end{eqnarray}
where the variances appearing in the denominators are determined by $\left< \Delta \boldsymbol{x}_n^2 \right>_0=\left< \left( \int_0^ndn'\,\overline{\boldsymbol{\eta}}_{n'} \right)^2 \right>=3k_BT\int_0^Ndn\int_0^Ndn'\, \partial_{n'}\mu_{st}(n,n')$ in the absence of force $\{ {\boldsymbol f}_{n} \}$.
Equation~(\ref{PDF_xn}) provides the number density of monomers at ${\boldsymbol x}$ given ${\boldsymbol x}_0$ through $\rho({\boldsymbol x}|{\boldsymbol x}_0)=\int dn \int d{\boldsymbol x}_n\, \delta ({\boldsymbol x}_n-{\boldsymbol x}) P({\boldsymbol x}_n|{\boldsymbol x}_0)\,\left[ \simeq \sum_n \int d{\boldsymbol x}_n\, \delta ({\boldsymbol x}_n-{\boldsymbol x}) P({\boldsymbol x}_n|{\boldsymbol x}_0) \right]$; thus, we encounter
\begin{eqnarray}
\rho({\boldsymbol x}|{\boldsymbol x}_0)
&=&
\int_0^N dn\,
\frac{1}{\sqrt{2\pi \left< \Delta \boldsymbol{x}_n^2 \right>_0 }}
\exp{} \Biggl(
\nonumber \\
&&
-\frac{\left( \Delta {\boldsymbol x}-\int_0^n dm\int_0^N dn'\,\mu_{st}(m,n'){\boldsymbol f}_{n'} \right)^2}{2\left< \Delta \boldsymbol{x}_n^2 \right>_0}
\Biggr)
\nonumber \\
\end{eqnarray}
with $\Delta {\boldsymbol x}\equiv {\boldsymbol x}-{\boldsymbol x}_0$.
Notably, $\rho({\boldsymbol x}|{\boldsymbol x}_0)$ is written with the kernel $\mu_{st}(m,n')$, $\left< \Delta \boldsymbol{x}_n^2 \right>_0$, and $\{ {\boldsymbol f}_{n} \}$.
The first two quantities are {\it a priori} obtained by supposing a mean uniform internal structure, whereas the force map $\{ {\boldsymbol f}_{n} \}$ is deduced from the heterogeneous distribution found in the experimental data.
A comparison of the analytical $\rho({\boldsymbol x}|{\boldsymbol x}_0)$ with the observation could be interesting for estimating a map of force acting on the chromatin.
The proposed analyses could enhance understanding of the configurations of the chromatin.
The details are left as future work.

\section{Concluding remarks}
\label{conclusion}

We have discussed the static GLE-like expression that describes an individual polymer configuration.
The static kernel $\mu_{st}(n,m)$ for the SA polymer
 corresponds to ``superdiffusive" in the language of anomalous diffusion. 
The formulation also covers the subdiffusion represented by a crumpled globule, although care is taken to note the difference (e.g., the sign of $1-2\nu$ in eq.~(\ref{noise_xn_corr_specific})). 

There are similarities and differences between the static and the dynamical GLEs.
As required in equilibrium statistical physics, the response function and the noise satisfy the static FRR with the monomer index variable; however, the form differs from that of the dynamical FRR appearing in the GLE. 
The remarkable differences are the translational or reversal symmetricity in the FRR and the sum of the fluctuation indices (eq.~(\ref{alpha_x_p}), (\ref{complementary_static})).
In addition, guided by the distinct form in the FRR, we considered the analogy with the stochastic energetics and the steady-state fluctuation theorem.

In the present article, we discussed a static formalism that enables us to deal with the nonlocal interaction.
This approach will hopefully contribute to the development of analyses of the distribution of the force acting on a single polymer, e.g., in a cell.

\section*{Acknowledgement}

Author thanks T. Sakaue for fruitful discussions and a critical reading.

\section*{Asymptotics in SA polymer}

We consider the asymptotic behaviors of eq.~(\ref{noise_xn_corr_specific}) without the mathematically rigorous arguments.
To be succinct, only the factor relevant to the asymptotics is extracted and transformed as
\begin{eqnarray}
\psi(n,n')
&\equiv&
\sum_{q= 1}^N \frac{1}{N} \left( \frac{q}{N} \right)^{1-2\nu} \cos{\left( \frac{\pi q(n-n')}{N} \right)}
\nonumber \\
&\simeq&
\int_{0}^1du\, u^{1-2\nu} \cos{\left[ \pi (n-n')u \right]}
\nonumber \\
&\simeq&
\left\{
\begin{array}{ll}
\int_{0}^1du\, u^{1-2\nu} \simeq 1  & |n-n'| \simeq 0
\\
|n-n'|^{2\nu-2} & |n-n'| \gg 1
\end{array}
\right..
\label{estimate_SA_corr}
\end{eqnarray}
where $u=q/N$ is introduced.
At the first step, $-\cos{\left( \pi q(n+n')/N \right)}$ in eq.~(\ref{noise_xn_corr_specific}) is ignored because it oscillates faster than $\cos{\left( \pi q(n-n')/N \right)}$ unless $n=n'=0$.
In the second line, $N$ is sufficiently large that the lower bound in the integral $1/N$ is replaced with $0$.

The two cases in the last line are not trivial, and we consider them as follows.

For $|n-n'|\simeq 0$, we use an approximation $\cos{\left[ \pi (n-n')u \right]} \simeq 1$ for $0\leq u \leq 1$.

For $|n-n'| \gg 1$, the integrand $u^{1-2\nu} \cos{\left[ \pi (n-n')u \right]}$ oscillates with respect to $u$, where the amplitude $u^{1-2\nu}$ becomes smaller as $u \rightarrow +\infty$.
We then apply an approximation: 
\begin{eqnarray}
\psi(n,n') &\simeq& \int_{0}^1du\, u^{1-2\nu} \cos{\left[ \pi (n-n')u \right]} 
\nonumber \\
&=& 
|n-n'|^{2\nu-2} \int_{0}^{1/2}ds\, s^{1-2\nu}\cos{\left[ \pi s \right]}
\nonumber \\
&& +|n-n'|^{2\nu-2}\sum_{j=1}^{|n-n'|-1} \int_{j-1/2}^{j+1/2}ds\, s^{1-2\nu}\cos{\left[ \pi s \right]}
\nonumber \\
&& 
+|n-n'|^{2\nu-2} \int_{|n-n'|-1/2}^{|n-n'|}ds\, s^{1-2\nu}\cos{\left[ \pi s \right]}
\label{powerlaw_cal1} \\
&\simeq& 
|n-n'|^{2\nu-2} \left(\frac{1}{2}\right)^{2-2\nu}
\nonumber \\
&&
+|n-n'|^{2\nu-2}\sum_{j=1}^{\infty} j^{1-2\nu} \int_{j-1/2}^{j+1/2}ds\, \cos{\left[ \pi s \right]}
\label{powerlaw_cal2} \\
&=& |n-n'|^{2\nu-2} \left[ \left(\frac{1}{2}\right)^{2-2\nu} -\sum_{j=1}^{\infty} (-1)^{j-1} j^{1-2\nu} \right]
\label{powerlaw_cal3} \\
&\simeq&
|n-n'|^{2\nu-2}.
\label{powerlaw_cal4} 
\end{eqnarray}
From the first line of eq.~(\ref{powerlaw_cal1}), we change the variable as $s=|n-n'|u$ and break the integral down depending on the signs of the cosine function $\cos{\left[ \pi s \right]}$.
In eq.~(\ref{powerlaw_cal2}), we approximate the upper bound in the sum by $+\infty$; in addition, the $j$-th values of a factor of the integrand $s^{1-2\nu}$ are represented by the midpoint value $j^{1-2\nu}$ over each integral range. The alternating series 
appearing in the bracket of eq.~(\ref{powerlaw_cal3}) is associated with the Riemann zeta function $\zeta(\lambda)$ as $\sum_{j=1}^{\infty} (-1)^{j-1}(1/j)^\lambda=(1-2^{1-\lambda})\zeta(\lambda)$.
Numerical computation indicates positive values $[(1/2)^{2-2\nu}-(1-2^{2-2\nu})\zeta(2\nu-1)]>0$ for $\nu=3/4, 3/5$ and the order of unity.
We then arrive at the other for $|n-n'| \gg 1$ in the last line.

\section*{Force fluctuations}

The dynamical GLE expression with eq.~(\ref{GLE_x}) is convenient for observing the displacement $\Delta x(t)$ when applying force $f(t)$.
In addition, there exists a paired expression that might be helpful in tracing the momentum transfer $\Delta p(t)=\int_0^tds\,f(s)$ when controlling the velocity $v(t)=dx(t)/dt$ (or the position $x(t)$).
Using the Laplace transform and organizing it, we rewrite eq.~(\ref{GLE_x}) as
\begin{eqnarray}
\frac{dp(t)}{dt}
=\int_0^tds\, \Gamma(t-s) v(s)+\eta^{(p)}(t),
\label{GLE_p}
\end{eqnarray}
where the Laplace transform is defined as $\hat{\phi}(\omega)\equiv \int_{0}^{\infty}dt\,\phi(t)e^{-\omega t}$ and the kernels retain the relation $\hat{\mu}(\omega)\hat{\Gamma}(\omega)=1$ in the Laplace domain.
Instead of eq.~(\ref{GLE_x_FDR}), the FRR is converted with $\Gamma(t)$ into
\begin{eqnarray}
\left< \eta^{(p)}(t)\eta^{(p)}(t') \right>=k_BT\Gamma(t-t').
\label{GLE_p_FDR}
\end{eqnarray}
Note that the superdiffusive kernel $\Gamma(t)$ creates $\alpha^{(p)}>1$ on the mean square ``momentum transfer" $\left< \Delta p(t)^2 \right>\sim t^{\alpha^{(p)}}$, whereas the subdiffusive $\mu(t)$ creates $\alpha^{(x)}<1$ on the ``mean square displacement" $\left< \Delta x(t)^2 \right>\sim t^{\alpha^{(x)}}$.

We here analogously consider the static counterparts to eq.~(\ref{GLE_xn}),\,(\ref{static_FDR_u}).
Applying $\Gamma_{st}(n,m)$ to eq.~(\ref{GLE_xn}), we obtain
\begin{eqnarray}
f_{n}
=
\int_0^Ndm\,\Gamma_{st}(n,m)\frac{dx_m}{dm} 
-\int_0^Ndm\,\Gamma_{st}(n,m)\overline{\eta}_m,
\label{fn_form_cal_01}
\end{eqnarray}
Comparing eq.~(\ref{fn_form_cal_01}) with eq.~(\ref{GLE_fn}), we have 
\begin{eqnarray}
\overline{\eta}_n^{(f)}
=
-\int_0^Ndm\, \Gamma_{st}(n,m)\overline{\eta}_{m}.
\label{eta_f}
\end{eqnarray}
As in the FRR (eq.~(\ref{static_FDR_u})), we relate $\overline{\eta}_n^{(f)}$ to $\Gamma_{st}(n,m)$.
To discover the FRR with $\overline{\eta}_n^{(f)}$ and $\Gamma_{st}(n,m)$, we assume a constant applied force (i.e., $F_q=const.$) because eq.~(\ref{GLE_fn}) should be maintained if either $f_n$ or $dx_{n'}/dn'$ is chosen as the controlled parameter. 
Using a solution to $X_q(t)$ in eq.~(\ref{x_mode}) under the time-independent $F_q$, we find that
\begin{eqnarray}
F_q h_{q,n}^{\dagger}
&=&
\left( \frac{\pi q}{N} \right)^{-1} k_q
\left( \frac{\pi q}{N} \right) X_q(t) h_{q,n}^{\dagger}
\nonumber \\
&&
-
\int_{-\infty}^tds\, \frac{k_q}{\gamma_q} Z_q(s) e^{-(t-s)(k_q/\gamma_q)}h_{q,n}^{\dagger}
\label{static_der}
\end{eqnarray}
such that the summation of the left side becomes $\sum_q F_q h_{q,n}^{\dagger}=f_n=-\partial T_n/\partial n$. 
Extracting the first term on the right-hand side and implementing a similar calculation around eq.~(\ref{mode_mu_F}),\footnote{
The first term on the right side of eq.~(\ref{static_der}) is transformed as
\begin{eqnarray}
&&
\sum_q 
\left( \frac{\pi q}{N} \right)^{-1} k_q
\left( \frac{\pi q}{N} \right) X_q(t) h_{q,n}^{\dagger}
\nonumber \\
&=&
\sum_{q,q',m} \left( \frac{\pi q}{N} \right)^{-1} k_q h_{q,n}^{\dagger}h_{q,m}^{(s)}
\left( \frac{\pi q'}{N} \right) X_{q'}h_{q',m}^{(s)\dagger} 
\nonumber \\
&=&
\sum_{m} 
\left( -
\sum_q
\left( \frac{\pi q}{N} \right)^{-1} k_q h_{q,n}^{\dagger}h_{q,m}^{(s)}
\right)
\left( -\sum_{q'} \left( \frac{\pi q'}{N} \right) X_{q'}h_{q',m}^{(s)\dagger}  \right)
\nonumber \\
\label{static_f_convol}
\end{eqnarray}
where $\partial_n x_n =-\sum_{q'} \left( \pi q'/N \right) X_{q'}(t) h_{q',n}^{(s)\dagger}$ in the last line.
Comparing eq.~(\ref{static_f_convol}) with $\int_0^N dm\, \Gamma_{st}(n,m)\partial_m x_m$ of eq.~(\ref{GLE_fn}), we arrive at eq.~(\ref{Gamma_st}).
}
we discover the kernel in the mode expression as
\begin{eqnarray}
\Gamma_{st}(n,n') 
=
-\sum_{q \geq 1} k_q \left( \frac{\pi q}{N} \right)^{-1} h_{q,n}^{\dagger}h_{q,n'}^{(s)}.
\label{Gamma_st}
\end{eqnarray}
The noise correlation for $\overline{\eta}_n^{(f)}$ is obtained by modifying the calculation in eq.~(\ref{noise_xn_corr}).
\begin{eqnarray}
\left< \overline{\eta}_n^{(f)} \overline{\eta}_{n'}^{(f)} \right> 
&=& \sum_{q\geq 1} \frac{c_qk_BTk_q}{N} h_{q,n'}^{\dagger} h_{q,n}^{\dagger}.
\label{noise_fn_corr}
\end{eqnarray}
Note that $h_{q,n}^{\dagger}$ or $h_{q,n}^{(s)\dagger}$ in eq.~(\ref{Gamma_st}),\,(\ref{noise_fn_corr}) substitutes for $h_{q,n}^{(s)\dagger}$ or $h_{q,n}^{\dagger}$ in eq.~(\ref{mu_st}),\,(\ref{noise_xn_corr}), respectively.
Comparing eq.~(\ref{noise_fn_corr}) with eq.~(\ref{Gamma_st}), we find an exact FRR relation:
\begin{eqnarray}
\left< \overline{\eta}_n^{(f)} \overline{\eta}_{n'}^{(f)} \right> 
= 
-k_BT \partial_{n'} \Gamma_{st}(n,n'),
\label{static_FDR_f}
\end{eqnarray}
where $\partial_{n'} h_{q,n'}^{(s)}=(\pi q/N)(c_q/N)h_{q,n'}^{\dagger}$.
In addition, comparing eq.~(\ref{noise_fn_corr}) with eq.~(\ref{noise_xn_corr}) leads to the static complementary relation eq.~(\ref{complementary_static}).

{\it --- Partition function ---}
\\
If the positions are controlled and the tension is observed, then
\begin{eqnarray}
Z_f[\{ x_n \}]
&=& \int DT_n\, Z_x[\{ T_n \}] \exp{\left(-\frac{\int_0^Ndn\,x_n\partial_nT_n}{k_BT}\right)}
\label{partion_Z_f}
\end{eqnarray}
substitutes for eq.~(\ref{partition_Zx}) as a partition function.
Accordingly, the thermodynamic potential can be defined as ${\cal F}_f\equiv-k_BT\log{Z_f}$.

As a typical tensile-deformed equilibrium, two end positions are pinned to $x_n=x_0\delta (n-\epsilon)+x_N\delta(n-N+\epsilon)$.
Note that the tension with $-\partial_n T_n|_{n=0}=\partial_n T_n|_{n=N}=-f$ acts on the chain ends and fluctuates such that the end-to-end distance $\Delta x$ can be controlled.
Analogously to eq.~(\ref{FRR_staric_x}), we obtain from ${\cal F}_f\equiv-k_BT\log{Z_f}$
\begin{eqnarray}
\frac{\partial \left<  f  \right>_{\Delta x}}{\partial (\Delta x)}
=
\frac{\left< \delta f^2\right>_{\Delta x}}{k_BT}.
\label{FRR_staric_f}
\end{eqnarray}

A remaining issue is to derive eq.~(\ref{static_FDR_f}) from ${\cal F}_f$, which is found by applying a similar approach to obtaining eq.~(\ref{FRR_partition_function}).

\section*{Derivation of eq.~(\ref{ratio_P_n})}

In this appendix, we derive eq.~(\ref{ratio_P_n}).
One of the forward configurational paths in the polymer $\{ x_0 \rightarrow x_1 \rightarrow \cdots  \rightarrow x_N \}$ is generated by the static GL eq.~(\ref{GLE_xn}) with a sequence of noise $\{ \overline{\eta}_0 \rightarrow \overline{\eta}_1 \rightarrow \cdots  \rightarrow \overline{\eta}_N \}$.
However, the reverse configurational path $\{ x_0^\dagger \rightarrow \cdots \rightarrow x_{N-1}^\dagger \rightarrow x_N^\dagger \}=\{ x_N \rightarrow \cdots \rightarrow x_1 \rightarrow x_0 \}$ should be generated by the same form as Langevin eq.~(\ref{GLE_xn}) as 
\begin{eqnarray}
\frac{dx_n^\dagger}{dn}
=\int_0^N dn'\, \mu_{st}(n,n')f_{n'}^\dagger +\overline{\eta}_n^{(R)},
\label{GL_xn_reverse}
\end{eqnarray}
but with a sequence of noise $\{ \overline{\eta}_0^{(R)} \rightarrow \overline{\eta}_1^{(R)} \rightarrow \cdots  \rightarrow \overline{\eta}_N^{(R)} \}$.
The controlled external parameters are analogously scheduled as $T_n^\dagger=T_{N-n}$ and also as $f_{n}^\dagger=-f_{N-n}$ obtained with $f_n=-\partial_n T_n$.

We here observe eq.~(\ref{GL_xn_reverse}) at the $m$-th monomer: $dx_m^\dagger/dm=\int_0^N dn'\, \mu_{st}(m,n')f_{n'}^\dagger +\overline{\eta}_m^{(R)}$.
The first term on the right-hand side is found to be
$\int_0^N dn'\, \mu_{st}(m,n')f_{n'}^\dagger=-\int_0^N dn'\, \mu_{st}(m,n')f_{N-n'}=\int_0^N dn'\, \mu_{st}(N-m,N-n')f_{N-n'}=\int_0^N dn'\, \mu_{st}(N-m,n')f_{n'}$ with anti-symmetricity $\mu_{st}(m,n')=-\mu_{st}(N-m,N-n')$ (see eq.~(\ref{mu_st})).
The left-hand side is sign-inverted as $dx_{m}^\dagger/dm=-dx_{m'}/dm'|_{m'=N-m}$.
Keeping these equations in mind, we organize eq.~(\ref{GLE_xn}) into
\begin{eqnarray}
-\frac{dx_{m'}}{dm'} \Biggr|_{N-m}
&=&
\int_0^N dn'\, \mu_{st}(N-m,n')f_{n'}
\nonumber \\
&&+\left( -2\int_0^N dn'\, \mu_{st}(N-m,n')f_{n'} -\overline{\eta}_{N-m} \right)
\nonumber \\
\label{GL_xn_reverse_mod}
\end{eqnarray}
such that the left-hand side and the first term on the right-hand side in eq.~(\ref{GL_xn_reverse}) correspond to those in eq.~(\ref{GL_xn_reverse_mod}).
Thus, the noise term in eq.~(\ref{GL_xn_reverse}) is found to be the last term in eq.~(\ref{GL_xn_reverse_mod}):
\begin{eqnarray}
\overline{\eta}_{N-n}^{(R)}
=
-2\int_0^N dn'\, \mu_{st}(n,n')f_{n'} -\overline{\eta}_{n}.
\end{eqnarray}
Note that $m=N-n$ has been substituted. 

In the same manner as eq.~(\ref{Pr_eta}), the probability of the reverse sequence of noise is obtained through
\begin{eqnarray}
{\cal P}[\overline{\eta}_{(\cdot)}^{(R)}]
&\sim&
\exp{\left(-\int_0^N dm\int_0^N dm'\,\frac{ (\partial_{m'}\mu_{st})^{-1}(m,m')}{2k_BT}\overline{\eta}_m^{(R)}\overline{\eta}_{m'}^{(R)} \right)}.
\nonumber \\
\end{eqnarray}
Focusing on the fact that $(\partial_{m'}\mu_{st})^{-1}(m,m')$ is unchanged under the variable transformation $(m,m')\rightarrow (N-m,N-m')$, we replace the integration variables with $n=N-m$ and $n'=N-m'$.
The ratio between the forward and reverse configurational paths is then considered as
\begin{eqnarray}
&&
\log{
\frac{{\cal P}[x_{(\cdot)}|x_0]}{{\cal P}[x_{(\cdot)^\dagger}|x_0^\dagger]}
}
\nonumber \\
&=&
\log{
\frac{{\cal P}[\eta_{(\cdot)}]}{{\cal P}[\eta_{(\cdot)}^{(R)}]}
}
\nonumber 
 \\
&=&
\int_0^N dn\int_0^N dn'\,\frac{ (\partial_{n'}\mu_{st})^{-1}(n,n')}{2k_BT}
\nonumber \\
&&
\times 
\Biggl[
4\int_0^N dm\, \mu_{st}(n,m)f_{m} \int_0^N dm'\, \mu_{st}(n',m')f_{m'}
\nonumber \\
&&
+
2\int_0^N dm\, \mu_{st}(n,m)f_{m} \overline{\eta}_{n}
+
2\int_0^N dm'\, \mu_{st}(n',m')f_{m'} \overline{\eta}_{n'}
\Biggr]
\nonumber \\
&=&
2\int_0^N dn\int_0^N dn'\,\frac{ (\partial_{n'}\mu_{st})^{-1}(n,n')}{k_BT}
\Biggl[
\frac{dx_{n'}}{dn'}-\overline{\eta}_{n'}
\Biggr]
\frac{dx_{n}}{dn},
\end{eqnarray}
where eq.~(\ref{GLE_xn}) is used from the third to the last equations.
Note that the Jacobian that appears in the conversion from $\{ x_{(\cdot)} \}$ to $\{ \eta_{(\cdot)} \}$ or from $\{ x_{(\cdot)^\dagger} \}$ to $\{ \eta_{(\cdot)}^{(R)} \}$ is eliminated.

\end{document}